\documentclass[12pt]{article}

\usepackage{epsfig,amssymb}

\makeatletter
\def\eqalign#1{\null\vcenter{\def\\{\cr}\openup\jot\m@th
  \ialign{\strut$\displaystyle{##}$\hfil&$\displaystyle{{}##}$\hfil
      \crcr#1\crcr}}\,}
\makeatother

\newcommand{\be}{\begin{equation}} 
\newcommand{\ee}{\end{equation}}
\newcommand{\beq}{\begin{eqnarray}}
\newcommand{\eeq}{\end{eqnarray}}
\newcommand{\bt}{\begin{theorem}}
\newcommand{\et}{\end{theorem}}
\newcommand{\bl}{\begin{lemma}}
\newcommand{\el}{\end{lemma}}
\newcommand{\bc}{\begin{corollary}}
\newcommand{\ec}{\end{corollary}}
\newcommand{\ba}{\begin{array}}
\newcommand{\ea}{\end{array}}

\newcommand{\la}{\label}
\newcommand{\ci}{\cite}

\newtheorem{theorem}{Theorem}
\newtheorem{lemma}[theorem]{LEMMA}
\newtheorem{corollary}[theorem]{COROLLARY}

\newcommand{\ti}{\tilde}
\newcommand{\de}{\delta}
\newcommand{\De}{\Delta}
\newcommand{\al}{\alpha}
\newcommand{\ga}{\gamma}
\newcommand{\Ga}{\Gamma}

\newcommand{\si}{\sigma}
\newcommand{\Si}{\Sigma}
\newcommand{\om}{\omega}
\newcommand{\Om}{\Omega}
\newcommand{\lb}{\lambda}
\newcommand{\ze}{\zeta}
\newcommand{\ka}{\varkappa}

\newcommand{\ep}{\varepsilon }

\newcommand{\A}{{\cal A}}

\newcommand{\bi}{\bibitem}

\newfont{\msbm}{msbm10 scaled\magstep1}%blackboardbold
\newfont{\msbms}{msbm7 scaled\magstep1} %blackboardbold   subscript

\newcommand{\bbr}{\mbox{$\mbox{\msbm R}$}}

\newcommand{\bbc}{\mbox{$\mbox{\msbm C}$}}

\begin{document}
\def\wt{\widetilde}
%\hfill {\small September 17, 2004}
\bigskip\bigskip\bigskip
\begin{center}
{\Large\bf 
Correlations of the characteristic polynomials in the 
Gaussian Unitary Ensemble or a singular Hankel determinant
}\\
\bigskip\bigskip\bigskip
I. V. Krasovsky\\
\bigskip 
\bigskip
Department of Mathematical Sciences,
Brunel University West London,\\
Uxbridge UB8 3PH, UK
\end{center}
\bigskip\bigskip
\noindent{\bf Abstract.} We obtain large $n$ asymptotics for 
products of powers of
the absolute values of the characteristic polynomials in the Gaussian Unitary
Ensemble of $n\times n$ matrices. Our results can also be interpreted as
asymptotics of the determinant of a Hankel matrix whose symbol is 
supported on the real
line and possesses power-like (Fisher-Hartwig) singularities. 
\bigskip\bigskip\bigskip

\section{Introduction}

In the present paper we compute large $n$ asymptotics for the
following averages of the characteristic polynomials over the Gaussian
Unitary Ensemble (GUE) of $n\times n$ matrices $H$:
\be
\langle \prod_{j=1}^m|\det(H-\mu_j)|^{2\al_j}\rangle_{\rm 
GUE}={D_n(\al_1,\dots,\al_m)\over D_n(0,\dots,0)},\qquad
\Re\al_j>-{1\over 2},\qquad j=1,\dots,m,\label{1}
\ee
where
\be
D_n(\al_1,\dots,\al_m)={1\over n!}\int_{-\infty}^\infty\cdots\int_{-\infty}^\infty
\prod_{i<j}{(x_i-x_j)^2}\prod_{k=1}^n{w(x_k)dx_k},\qquad
w(x)=\prod_{j=1}^m|x-\mu_j|^{2\al_j}e^{-x^2},
\ee
and all $\mu_j$ are distinct.
The condition $\Re\al_j>-{1\over 2}$ guarantees convergence.

It is easy to verify that
\be
D_n(\al_1,\dots,\al_m)=\det(M_{i+j})_{i,j=0}^{n-1},\qquad 
M_k=\int_{-\infty}^\infty x^k w(x)dx,
\ee
that is, $D_n(\al_1,\dots,\al_m)$ is a Hankel determinant with symbol $w(x)$.

The determinant $D_n(0,\dots,0)$ is an $n$-dimensional Selberg integral
\ci{Mehta}, it can be evaluated exactly, and its large-$n$ asymptotics
readily can be written (cf. \ci{Warc}):
\be\label{D0}
D_n(0,\dots,0)=(2\pi)^{n/2}2^{-n^2/2}\prod_{j=1}^{n-1}{j!}=
(2\pi)^n(n/2)^{n^2/2}n^{-1/12}e^{-(3/4)n^2+\ze'(-1)}(1+O({1/n})),
\ee
where $\ze'(x)$ is the derivative of Riemann's zeta-function.

The average (\ref{1}) is related to several
interesting questions. We mention a conjectured relation of (\ref{1})
for $m=1$
and such averages over other ensembles of random matrices to the mean
values $\{1/T\}\int_0^T|\ze(1/2+it)|^{2\al}dt$ of the zeta-function on the
critical line \ci{K,KLR,BH}. Also note that $D_n(\al_1,\dots,\al_m)$ is a Hankel
determinant with symbol on $\bbr$ and Fisher-Hartwig
singularities at $x=\mu_j$. Although asymptotics of Toeplitz determinants with
general Fisher-Hartwig symbols have by now largely been determined 
(at least the leading terms thereof, see,
e.g., \ci{Forr}), the Hankel case is still open. 

For connections of
(\ref{1}) to the one-dimensional gas of impenetrable bosons and other
physical problems, see \ci{Forr, FFGW, Fyod, Ga} and below.
(see also \ci{BDS}, \ci{BS}, \ci{Ga}, \ci{MN}, \ci{SF}, \ci{Va} 
regarding subleading asymptotic terms,
negative moments and other random matrix ensembles).

The asymptotics of averages (\ref{1}) for $\al_j$ positive integers
have been found by Br\'ezin and Hikami \ci{BH}, Forrester and Frankel \ci{Forr},
Garoni \ci{G}.
For such $\al_j$'s, the expression (\ref{1}) can be reduced to the 
Hermite polynomials
and their derivatives at the points $\mu_j$.
(In the framework of the present paper, this can be shown with the help of 
Christoffel's formula (e.g.,\ci{Sz}, p.29), cf. \ci{BDS})
It is not, however, the case for noninteger $\al_j$'s.
We prove 

\noindent
{\bf Theorem 1} {\it 
Fix $\lb_j$ and $\al_j$ such that $\lb_j\in(-1,1)$, 
$\lb_j\neq\lb_k$, for $j\neq k$, and $\Re\al_j>-1/2$, $j,k=1,\dots,m$. 
Then, as $n\to\infty$,
\be\la{a}\eqalign{
\langle \prod_{j=1}^m|\det(H-\lb_j\sqrt{2n})|^{2\al_j}\rangle_{\rm GUE}=
\prod_{j=1}^m \left[ C(\al_j)(1-\lb_j^2)^{\al_j^2/2}
\left({n\over 2}\right)^{\al_j n+\al_j^2}\exp\left\{(2\lb_j^2-1)\al_j n
\right\}\right]\times\\
\prod_{1\le i< j\le m}(2|\lb_i-\lb_j|)^{-2\al_i\al_j}
\left[1+O\left({\ln n\over n}\right)\right],}\label{as}
\ee
where  
\be\la{c}
C(\al)=\Gamma(\al+1/2)^{-2\al}\exp\left(2\int_0^\al
  \ln\Gamma(s+1/2)ds+
\al^2\right)=2^{2\al^2}{G(\al+1)^2\over G(2\al+1)},
\ee
and $G(z)$ is Barnes' $G$-function. In (\ref{a}), the values of
the roots positive for real $\al_j$ (and their analytic extension for
complex $\al_j$) are taken.
} 

With increasing effort, one can compute an arbitrary number of terms
in these asymptotics (and verify a heuristic calculation of them by
Gangardt \ci{Ga} for $m=1$).

The expression (\ref{as}) without the error term was conjectured 
in \ci{Forr}, Conjecture 4, as an extension off integer values of $\al_j$.

The asymptotics for the Hankel determinant $D_n(\al_1,\dots,\al_2)$ 
are now just a combination of (\ref{1}), (\ref{D0}), and (\ref{as}).
The corresponding asymptotics for a Toeplitz determinant with symbol 
(on the unit circle) having
such type of singularities (multiplied by a rather general ``regular'' function) 
were found by Widom \ci{W}. The multiplicative constant
in the asymptotics involves the same combination of $G$ functions as here.
Note that averages of type (\ref{1}) for the Circular Unitary Ensemble 
are such singular Toeplitz determinants. The conjectured relation to Riemann's
zeta-function averages was first formulated by Keating and Snaith
\ci{K} for that ensemble.

Let $m=1$. Since the density of the scaled eigenvalues in the GUE is given by
$\psi(\lb)=(2/\pi)\sqrt{1-\lb^2}$, $-1<\lb<1$ (the Wigner semicircle law),
we have that $D_n(\al)$ is proportional to $\psi^{\al^2}$. Similar
dependence on zero density as well as presence of the factor 
$G(\al+1)^2/ G(2\al+1)$ is conjectured for the large 
$T$ asymptotics of the zeta-function average mentioned above.

For $\al=1/2$, $m=1$, Theorem 1 gives an analogue for the GUE of the average of 
the absolute value of the characteristic polynomial found by Fyodorov
for the Gaussian Orthogonal Ensemble \ci{Fyod} (in the Unitary Ensemble, his
method would correspond to the case $\al=1$).

The condition $\lb\in(-1,1)$ means that $\lb$ is inside 
the bulk of the scaled spectrum of $H$ (in other words, inside the
support of the equilibrium measure for the GUE). The case
$\lb\in\bbc\setminus [-1,1]$ corresponds to a Hankel determinant with
symbol ``regular'' on the support of the equilibrium measure, and
the average (\ref{1}) is then found as a particular case of a result
by Johansson \ci{J}, although some nonessential assumptions in \ci{J}
need to be relaxed. (This result gives a Hankel analogue of the Szeg\H o asymptotics
for Toeplitz determinants with a regular symbol.
The need to consider the equilibrium measure is an additional difficulty 
in comparison with the Toeplitz case.)
For real $\lb$, $\al$, $m=1$, we obtain from \ci{J} as $n\to\infty$:
\be\eqalign{
 \langle |\det(H-\lb\sqrt{2n})|^{2\al}\rangle_{\rm GUE}=
(2n)^{\al n}(\lb^2-1)^{-\al^2}
\left({|\lb|+\sqrt{\lb^2-1}\over 2}\right)^{2\al(n+\al)}\times\\
\exp\left\{2\al n \left(\lb^2-|\lb|\sqrt{\lb^2-1}-{1\over 2}\right)
\right\}[1+o(1)],\qquad \lb\in\bbr\setminus[-1,1],\qquad\al>-1/2.}\label{as2}
\ee
This (including the error terms) and
the asymptotics for the border case of $\lb$ close to $\pm 1$ could also be
computed by our methods. 
It should also be possible to generalize Theorem 1 replacing
$x^2$ in $w(\sqrt{2n}x)$ by a more general function $V(x)$. 
Thus it should be possible to verify Conjecture 8 of Forrester and Frankel \ci{Forr}.

The one-body density matrix $\rho(\lb_1,\lb_2)$ 
for the ground state wave function
of the one-dimensional gas of 
impenetrable bosons equals \ci{FFGW}, up to a simple factor, 
to $D_n(1/2,1/2)$ (i.e., the corresponding $w(x)=|x-\lb_1||x-\lb_2|e^{-x^2}$),
where $n$ represents the number of particles.
The eigenvalues of the operator with kernel  $\rho(\lb_1,\lb_2)$ are interpreted 
as occupation numbers of effective single-particle states of the Bose system. 
To prove the expected 
asymptotic $\sqrt{n}$-proportionality of the leading occupation numbers
in this one-dimensional system
(note that Bose-Einstein condensation, which is not expected in one dimension,
would correspond to the $n$-proportionality),
we need to analyze asymptotics of $D_n(1/2,1/2)$ for all values of $\lb_1$, $\lb_2$.
Theorem 1 gives the desired asymptotics (conjectured in \ci{FFGW}, eq.(105))
for $-1<\lb_1<\lb_2<1$, but we still need to analyze, in particular, cases 
of $\lb_1$, $\lb_2$ approaching each other and the points $\pm 1$. 
This will be the subject of a subsequent publication.

We prove Theorem 1 using Riemann-Hilbert problem methods (see
\ci{Dbook} for an introduction). Our approach is as follows. Consider
the polynomials $p_n(x)=\ka_n x^n+\cdots$ orthonormal w.r.t. to the
weight $w(x)=\prod_{j=1}^m|x-\lb_j\sqrt{2n}|^{2\al_j}\exp(-x^2)$:
\be\la{OP}
\int_{-\infty}^\infty p_j(x)p_k(x)w(x)dx=\de_{j\,k},\qquad
j,k=0,1,\dots
\ee
Then, 
\be
D_n(\al_1,\dots,\al_m)=\prod_{j=0}^{n-1}\ka_j^{-2},\la{Dchi}
\ee
by a well-known 
formula \ci{Sz,Dbook} (for real $\al_j$; the case of complex $\al_j$ will 
be explained below).
The asymptotics of the polynomials $p_n(x)$ 
(for $\lb=0$ and $m=1$) as $n\to\infty$ were recently analyzed by Kuijlaars 
and Vanlessen 
in \ci{KV} by a Riemann-Hilbert problem approach. Using \ci{KV}, one
can obtain any number of asymptotic terms for $p_n(x)$ and $\ka_n$.

Thus, knowing $\ka_j$, $j\to\infty$, we can shed light on the
asymptotics of $D_n(\al_1,\dots,\al_m)$. However, since the product of the first
$\ka_j$, $j=1,2,\dots$ remains unknown we are bound to lose at least
the constant (in $n$) factor in such obtained asymptotics of
$D_n$ (cf. \ci{KVA}). This problem is avoided here by deriving an
identity for $(d/d\al_\nu)\ln D_n(\al_1,\dots,\al_m)$, $\nu=1,\dots,m$. 
(cf. Deift's formula \ci{Ddiff}
and differential identities of \ci{DIZ}), an idea which has
proven useful in similar situations \ci{EM,BI,Kr,DIKZ,DIK2}. It turns out that
the particular structure of the weight $w(x)$ allows us to write 
the above logarithmic derivative {\it only} in terms of $p_n(z)$,
$\int_{-\infty}^\infty p_n(x)w(x)dx/(x-z)$, and similar expressions with index $n-1$,
which are precisely the
quantities whose asymptotics (uniform for $\al$ in a bounded set) 
follow from the Riemann-Hilbert problem for $p_n(z)$.
A fact that considerably simplifies calculations is that these quantities only need
to be evaluated at a finite number of points $z$
($\mu_j$ and infinity); in this sense the identity we obtain 
is ``local''. (As discussed in Section 3, a ``nonlocal'' identity involving integrals 
with $p_n(x)$, $p_{n-1}(x)$ exists for general weights).
The identity is found in Section 3. 
Setting in it all $\al_j=0$, $\nu=1$, and integrating over $\al_1$ from
zero to some $\al_1$, we obtain the asymptotics for 
$\ln (D_n(\al_1,0,\dots,0)/D_n(0,\dots,0))$.
Now fixing in the identity $\al_1$, setting $\al_2=\cdots=\al_m=0$, $\nu=2$, and
integrating over $\al_2$ from zero to some $\al_2$, we obtain
$\ln (D_n(\al_1,\al_2,0,\dots,0)/D_n(\al_1,0,\dots,0))$.
Continuing this procedure, we prove the theorem by induction.
Note that instead of zero, we could have used any positive integer $\al$ 
(where, as noted above, the asymptotics of $D_n$ are known) as a starting point 
for the integration.

\section{Riemann-Hilbert problem for $p_n(z)$}

Consider the following Riemann-Hilbert problem for a 
$2\times 2$ matrix valued function $Y(z)$ and the weight
$w(x)=\prod_{j=1}^m|x-\mu_j|^{2\al_j}\exp(-x^2)$, $\Re\al_j>-1/2$:
\begin{enumerate}
    \item[(a)]
        $Y(z)$ is  analytic for $z\in\bbc \setminus\bbr$.
    \item[(b)] 
Let $x\in\bbr\setminus\cup_{j=1}^m\{\mu_j\}$.
$Y$ has continuous boundary values
$Y_{+}(x)$ as $z$ approaches $x$ from
above, and $Y_{-}(x)$, from below. 
They are related by the jump condition
\begin{equation}\label{RHPYb}
            Y_+(x) = Y_-(x)
            \pmatrix{
                1 & w(x) \cr
                0 & 1},
            \qquad\mbox{$x\in\bbr\setminus\cup_{j=1}^m\{\mu_j\}$.}
        \end{equation}
    \item[(c)]
        $Y(z)$ has the following asymptotic behavior at infinity:
        \begin{equation} \label{RHPYc}
            Y(z) = \left(I+ O \left( \frac{1}{z} \right)\right)
            \pmatrix{
                z^{n} & 0 \cr
                0 & z^{-n}}, \qquad \mbox{as $z\to\infty$.}
        \end{equation}
\item[(d)]
The matrix of 
$Y(z)$ is $O(1)$ for $\Re\al_j\ge 0$,
and $\pmatrix{ O(1) & O(|z-\mu_j|^{2\al_j})\cr O(1) & O(|z-\mu_j|^{2\al_j})}$
for $\Re\al_j<0$ as $z\to\mu_j$, $j=1,\dots,m$,
$z\in\bbc\setminus\bbr$.

\end{enumerate}
(Here and below $O(a)$ stands for $O(|a|)$.)

It is easy to verify that,
provided the system of orthogonal polynomials $p_k(z)=\ka_k z^k+\cdots$, $\ka_k\neq 0$,
$k=0,1,\dots$, satisfying (\ref{OP}) exists, this problem has a solution given by
the function:
\begin{equation} \label{RHM}
    Y(z) =
    \pmatrix{
\ka_n^{-1}p_n(z) & 
\ka_n^{-1}\int_{-\infty}^{\infty}{p_n(\xi)\over \xi-z}
{w(\xi)d\xi \over 2\pi i } \cr
-2\pi i\ka_{n-1}p_{n-1}(z) & 
-\ka_{n-1}\int_{-\infty}^{\infty}{p_{n-1}(\xi)\over \xi-z} w(\xi)d\xi 
    }.
\end{equation}
For general weights
this fact was noticed by Fokas, Its, and Kitaev \ci{fik} and, in conjunction 
with the steepest
descent method of Deift and Zhou \ci{dz} for asymptotic analysis of matrix
Riemann-Hilbert 
problems, has allowed in recent years to solve many previously
unaccessible asymptotic problems for orthogonal polynomials 
(see \cite{Dbook} for an introduction and bibliography of earlier works, and 
\cite{KV,V,KVA,KM,BK,JBD,Kr,DIKZ}).

Note, in particular, that $\det Y(z)=1$. (From the conditions on $Y(z)$ it 
follows that $\det Y(z)$ is analytic across the real axis, has all singularities 
removable, and tends to the identity as $z\to\infty$. 
It is then identically $1$ by the Liouville theorem.) The solution is unique. Indeed,
if there is another solution $\wt Y(z)$, we easily obtain by the Liouville theorem
that $Y(z) \wt Y(z)^{-1}\equiv 1$.

The case of the weight $w(x)=|x|^{2\al}\exp(-x^2)$, $\al>-1/2$,
was considered in \cite{KV} (the argument is
straightforward to generalize to our case). Namely, \ci{KV} gives us a ready 
procedure to calculate asymptotics of $Y(z)$ in this case (by applying
a series of transformations to the problem (a)--(d)). By (\ref{RHM}),
these results are then interpreted as asymptotics of the polynomials $p_n(z)$
and their Cauchy transforms.

In the next section we shall derive an expression for 
$(d/d\al_j)\ln D_n(\al_1,\dots,\al_m)$
in terms of the matrix elements of (\ref{RHM}), and 
in the section after that compute the asymptotics of $Y(z)$.

The existence of the system of orthogonal polynomials $p_k(z)=\ka_k z^k+\cdots$ 
satisfying (\ref{OP})
with nonzero leading coefficients $\ka_k$ for $\al_j$ real, $\al_j>-1/2$, is a 
classical fact. Moreover, the coefficients $\ka_k^2=D_k/D_{k+1}$ are regular functions 
of all $\al_j$
(as follows from the determinantal representation for $D_k$ \ci{Sz, Dbook}). 
For all complex $\al_j$ 
in any fixed closed bounded set of the half-plane $\Re\al_j>-1/2$, $j=1,\dots,m$
(denote the set of such $m$-tuples $\{\al_j\}_{j=1}^m$ by $\wt\Om$),
we shall prove below the existence of a solution to
the Riemann-Hilbert problem for all $n$ larger than some $n_0>0$. Its asymptotics
will be explicitly 
constructed. We shall also see that the coefficients $\ka_k$, $k>n_0$ are nonzero 
and finite for all such $\al_j$. For $k\le n_0$, the coefficients $\ka_k^2$ are regular
and nonzero (as follows from the determinantal representation) and $D_k$ are nonzero
outside of a possible subset $\hat\Om$ of $\wt\Om$. 
When all $\al_j$ except for $\al_{j_0}$
(for some $j_0$) are fixed provided only $\{\al_j\}_{j=1}^m\in\wt\Om$, 
we denote
the set of values of $\al_{j_0}$ such that $\{\al_j\}_{j=1}^m\in\hat\Om$ by 
$\Om(\al_1,\dots,\al_{j_0-1},*,\al_{j_0+1},\dots,\al_m)$. 
As a consequence of the 
determinantal representation, the system of the orthogonal polynomials exists, and
the formula (\ref{Dchi}) holds for $\al_{j_0}$ in the $j_0$-component of 
$\wt\Om$, outside the 
set $\Om(\al_1,\dots,\al_{j_0-1},*,\al_{j_0+1},\dots,\al_m)$. Throughout Section 3 
(and, hence, in the differential identity 
obtained there) we assume that $\{\al_j\}_{j=1}^m\notin\hat\Om$. 
(There is no such condition on $\al_j$ in Section 4.) This provides in Section 5
a proof of Theorem 1 for $\{\al_j\}_{j=1}^m$ outside the set $\hat\Om$. 
However, as we shall see
below, the error term in the asymptotics of $D_n$ is uniform for {\it all} 
$\{\al_j\}_{j=1}^m\in\wt\Om$.
Theorem 1 will follow then by continuity.

\section{Differential identity}
Throughout this section, we consider $n$ a {\it fixed} positive integer and
$\{\al_j\}_{j=1}^m\in \wt\Om\setminus\hat\Om$ (see Section 2).
Let us fix some $\nu$ from $1$ to $m$, and denote $\al_\nu=\al$. 
The orthogonality property of the polynomials $p_n(x)=\ka_n
x^n+\cdots$ implies that
\be
\int_{-\infty}^\infty p_k(x)x^j w(x)dx={\de_{jk}\over\ka_j},\qquad 
j=0,1,\dots,k,\quad k=0,1,2,\dots
\ee
Using (\ref{Dchi}) and this relation, we have
\be\la{D1}\eqalign{
{d\over d\al}\ln D_n(\al_1,\dots,\al_m)={d\over d\al}\ln 
\prod_{j=0}^{n-1}\ka_j^{-2}=
-2\sum_{j=0}^{n-1}{\ka'_{j,\al}\over \ka_j}=
-2\sum_{j=0}^{n-1}
\int_{-\infty}^\infty p_j(x)p'_{j,\al}(x)w(x)dx=\\
-\int_{-\infty}^\infty \left(\sum_{j=0}^{n-1}p^2_j(x)\right)'_\al
w(x)dx.}
\ee
Here the prime and the lower index $\al$ stand for the derivative w.r.t. $\al$. 
Below we also use 
derivatives w.r.t. $x$ denoting them with the prime and the lower index $x$.

Note that $p_n(x)$ are analytic functions of $\al_j$
as follows, e.g., from their representation as a determinant and regularity
of the leading coefficients for 
$\{\al_j\}_{j=1}^m\in \wt\Om\setminus\hat\Om$.

By the well-known Christoffel-Darboux formula (e.g., \ci{Sz}),
\be\la{CD}
\sum_{j=0}^{n-1}p^2_j(x)=b_{n-1}(p'_{n,x}(x)p_{n-1}(x)-p_n(x)p'_{n-1,x}(x)),
\ee
where $b_j$ are coefficients in the recurrence relation for our
polynomials:
\[
b_{j-1}p_{j-1}(x)+(a_j-x)p_j(x)+b_j p_{j+1}(x)=0,\qquad j=1,2,\dots
\]
Let us fix the notation for the 3 leading coefficients of the polynomials
$p_n(z)$ as follows:
\[
p_n(z)\ka_n^{-1}=z^n+\beta_n z^{n-1}+\ga_n z^{n-2}+\cdots
\]
Comparing the coefficients at the powers $z^{n+1}$, $z^n$, and
$z^{n-1}$ in the recurrence relation, we obtain the following identities
we shall use later on:
\be\la{coeffid}
b_j={\ka_j\over \ka_{j+1}},\qquad 
a_j=\beta_j-\beta_{j+1},\qquad
\left(\ka_{j-1}\over \ka_j\right)^2=\ga_j-\ga_{j+1}
-\beta_j^2+\beta_j\beta_{j+1}.
\ee 

Substituting (\ref{CD}) into (\ref{D1}), differentiating it
w.r.t. $\al$, using the orthogonality and the above expression for
$b_{n-1}$, we obtain:
\be\la{D2}\eqalign{
{d\over d\al}\ln D_n(\al_1,\dots,\al_m)=
-n{\ka'_{n-1,\al}\over\ka_{n-1}}+{\ka_{n-1}\over\ka_n}(J_1-J_2),\\
J_1=\int_{-\infty}^\infty p'_{n,\al}(x)p'_{n-1,x}(x)w(x)dx,
\qquad
J_2=\int_{-\infty}^\infty p'_{n,x}(x)p'_{n-1,\al}(x)w(x)dx.}
\ee
 
Note that since (\ref{D2}) contains polynomials with indices $n-1$ and $n$ only,
it is already an identity one could use to obtain an asymptotic expression for 
the logarithmic derivative. This identity is valid for a general weight $w(x)$.
However, to obtain an asymptotic expression for $(d/d\al)\ln D_n$, one would 
have to integrate asymptotics of the polynomials over the whole real line,
which would make the calculation rather cumbersome. 
We shall now see that the particular structure of our weight $w(x)$ allows us
to simplify (\ref{D2}) considerably and reduce it to an identity
involving only some polynomial coefficients and the
values of polynomials and their Cauchy transforms at the points 
$\mu_j$. 

Let us  evaluate $J_1$. 
Choose points $c_k$, $d_k$ according to the inequalities:
\be
-\infty=c_1<\mu_1<d_1=c_2<\mu_2<d_2=c_3<\cdots<d_{m-1}=c_m<\mu_m<d_m=\infty
\ee
For a small $\ep>0$,
split the integral into $3m$ ones over the intervals 
$(c_k,\mu_k-\ep)$, $(\mu_k-\ep,\mu_k+\ep)$,
$(\mu_k+\ep,d_k)$. The middle ones can be written in the form
\[
\int_{\mu_k-\ep}^{\mu_k+\ep}|x-\mu_k|^{2\al_k}f(x)dx,
\]
where $f(x)$ is a smooth function on $(\mu_k-\ep,\mu_k+\ep)$. 
Hence, this integral is of 
order $O(\ep^{2\al_k+1})$, small for $\Re\al_k>-1/2$. For the same reason
\[
f(x)|x-\mu_k|^{2\al_k}|^{\mu_k+\ep}_{\mu_k-\ep}=
(f(\mu_k+\ep)-f(\mu_k-\ep))\ep^{2\al_k}=O(\ep^{2\al_k+1}).
\]
Now we can integrate each of the remaining integrals
\[
\int_{c_k}^{\mu_k-\ep} p'_{n,\al}(x)p'_{n-1,x}(x)w(x)dx+
\int_{\mu_k+\ep}^{d_k} p'_{n,\al}(x)p'_{n-1,x}(x)w(x)dx
\]
by parts and then take the limit $\ep\to 0$ to obtain $J_1$.
Since $w(x)$ decreases as $e^{-x^2}$ at
infinity and
\be
w(x)'_{x}=\left(\sum_{j=1}^m{2\al_j\over x-\mu_j}-2x\right)w(x),
\ee
integration by parts gives (we again use the orthogonality to write
$\int p_{n-1}p''_{n,x,\al}wdx=n\ka'_{n,\al}/\ka_{n-1}$)
\be\la{J1}
J_1=-n{\ka'_{n,\al}\over\ka_{n-1}}-2\sum_{j=1}^m\al_j
\int_{-\infty}^\infty p_{n-1}(x){ p'_{n,\al}(x)\over x-\mu_j}w(x)dx+
2\int_{-\infty}^\infty p_{n-1}(x)p'_{n,\al}(x)x w(x)dx,
\ee
where the first $m$ integrals are taken in the principal value sense.
Note that they converge for $\Re\al_j>-1/2$. This can be seen directly as follows.
Take the $j$'s one and write it in the form
\[
\int_{-\infty}^\infty\frac{|x-\mu_j|^{2\al}}{x-\mu_j}f(x)dx=\lim_{\ep\to 0}
\left(
\int_{-\infty}^{\mu_j-\ep}-(\mu_j-x)^{2\al_j-1}f(x)dx+
\int_{\mu_j+\ep}^\infty(x-\mu_j)^{2\al_j-1}f(x)dx\right),
\]
where $f(x)$ is a smooth function at $x=\mu_j$. Change the variables $y=\mu_j-x$ in the first,
and $y=x-\mu_j$ in the second integral on the r.h.s. Then the sum reduces to
\[
\int_\ep^\infty y^{2\al_j-1}(f(\mu_j+y)-f(\mu_j-y))dy.
\]
The integrand is of order $O(y^{2\al_j})$ for small $y$, and therefore the integral converges 
as $\ep\to 0$.

We now simplify the integral. Adding and
subtracting $p'_{n,\al}(\mu_j)$ gives:
\[\eqalign{
\int_{-\infty}^\infty p_{n-1}(x){ p'_{n,\al}(x)\over x-\mu_j}w(x)dx=\\
\int_{-\infty}^\infty p_{n-1}(x){ p'_{n,\al}(x)-p'_{n,\al}(\mu_j)\over 
x-\mu_j}w(x)dx+
p'_{n,\al}(\mu_j)\int_{-\infty}^\infty {p_{n-1}(x)w(x) \over x-\mu_j}dx.}
\]
The ratio in the first integral on the r.h.s. 
is obviously a polynomial of degree
$n-1$ in $x-\mu_j$ with the leading coefficient $\ka'_{n,\al}$.
Therefore, by the orthogonality,
\[
\int_{-\infty}^\infty p_{n-1}(x){ p'_{n,\al}(x)-p'_{n,\al}(\mu_j)\over 
x-\mu_j}w(x)dx=
{\ka'_{n,\al}\over \ka_{n-1}}.
\]
The last integral in the expression (\ref{J1}) for $J_1$ can be 
rewritten as follows:
\[
\int_{-\infty}^\infty
p_{n-1}(x)(\ka'_{n,\al}x^{n+1}+(\ka_n\beta_n)'_\al
x^n+(\ka_n\ga_n)'_\al x^{n-1}+\cdots)w(x)dx.
\]
Expanding here $x^{n+1}$ and $x^n$ in terms of the polynomials $p_j(x)$,
we obtain by the orthogonality and (\ref{coeffid}) that this integral
equals 
\be
{\ka_{n}\over \ka_{n-1}}\left[
{\ka'_{n,\al}\over \ka_n}
\left(\ka_{n-1}\over \ka_n\right)^2
+\ga'_{n,\al}-\beta_n\beta'_{n,\al}\right].
\ee
We, therefore, finally obtain
\be\la{J1e}\eqalign{
J_1=-(n+2\sum_{j=1}^m\al_j){\ka'_{n,\al}\over\ka_{n-1}}
-2\sum_{j=1}^m\al_j p'_{n,\al}(\mu_j)\int_{-\infty}^\infty {p_{n-1}(x)w(x)\over x-\mu_j}dx+\\
2{\ka_{n}\over \ka_{n-1}}\left[
{\ka'_{n,\al}\over \ka_n}
\left(\ka_{n-1}\over \ka_n\right)^2
+\ga'_{n,\al}-\beta_n\beta'_{n,\al}\right].}
\ee

The analysis of $J_2$ is similar (and simpler). We obtain that
\be\la{J2e}
J_2=2{\ka'_{n-1,\al}\over \ka_n}-
2\sum_{j=1}^m\al_j p'_{n-1,\al}(\mu_j)\int_{-\infty}^\infty {p_{n}(x)w(x)\over
  x-\mu_j}dx,
\ee
again with the principal value integral.

Note that at $\al_j=0$ in the above two equations
the terms multiples of $\al_j$ disappear.

Thus, by (\ref{D2}), (\ref{J1e}), (\ref{J2e}), and (\ref{RHM}), we can now write our
identity in terms of the matrix elements of $Y(z)$.
Some care is needed when comparing the Cauchy transforms in the second column of $Y(z)$
with the principal value integrals in $J_1$ and $J_2$: in particular, note that
$Y_{12}(z)$ and $Y_{22}(z)$ are unbounded at $z=\mu_j$ if $\Re\al_j<0$.

Consider $-1/2<\Re\al_j\le 0$, $\al_j\neq 0$.
The contribution to the matrix elements $Y_{12}(z)$ and $Y_{22}(z)$
containing a singular part has the form
\be\la{i1}
\int_{c_j}^{d_j}\frac{|x-\mu_j|^{2\al_j}}{x-z}f(x)dx
\ee
with a function $f(x)$ analytic in a neighborhood of $x=\mu_j$
(analytic continuation off the real axis).
On the other hand, the corresponding term in (\ref{J1e}) or (\ref{J2e}) has the form
\be\la{i2}
\int_{c_j}^{d_j}\frac{|x-\mu_j|^{2\al_j}}{x-\mu_j}f(x)dx
\ee
with the same $f(x)$ and the integral taken in the sense of the principal value.
The difference of (\ref{i1}) and (\ref{i2}) can be written as the principal 
value integral
\be\la{pv}
(z-\mu_j)\lim_{\ep\to 0}
\left(\int_{c_j}^{\mu_j-\ep}\frac{|x-\mu_j|^{2\al_j}}{(x-z)(x-\mu_j)}f(x)dx+
\int_{\mu_j+\ep}^{d_j}\frac{|x-\mu_j|^{2\al_j}}{(x-z)(x-\mu_j)}f(x)dx\right).
\ee
Consider
\be\la{iC}
\int_C \frac{(x-\mu_j)^{2\al_j}}{(x-z)(x-\mu_j)}f(x)dx,
\ee
where the contour $C$ is shown in Figure 1.

\begin{figure}
\centerline{\psfig{file=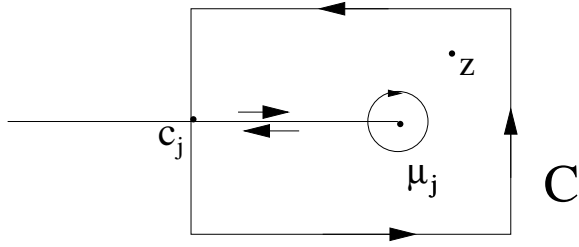,width=3.0in,angle=0}}
\vspace{0cm}
\caption{
Integration contour for (\ref{iC}).}
\label{fig0}
\end{figure}

On the one hand, this integral
equals its residue at $x=z$
\be
2\pi i (z-\mu_j)^{2\al_j-1}f(z).
\ee
On the other hand, we can rewrite the integral as a sum of parts along 
the real axes, a circle
(of radius $\ep$) around $x=\mu_j$, 
and the rest of the contour (denote this rest $C_0$):
\be\eqalign{
\int_C \frac{(x-\mu_j)^{2\al_j}}{(x-z)(x-\mu_j)}f(x)dx=
(e^{2\pi i\al_j}-e^{-2\pi i\al_j})
\int_{c_j}^{\mu_j-\ep}\frac{|x-\mu_j|^{2\al_j}}
{(x-z)(x-\mu_j)}f(x)dx+\\
f(\mu_j){\ep^{2\al_j}\over \mu_j-z}{e^{-2\pi i\al_j}-
e^{2\pi i\al_j}\over 2\al_j}+{O(\ep^{2\al_j+1})\over z-\mu_j}+F_1(z),}
\ee
where $F_1(z)$ is regular as $z\to\mu_j$ and given by the integral over $C_0$.
The last 2 equations give the first integral in (\ref{pv}):
\[
\eqalign{
\int_{c_j}^{\mu_j-\ep}\frac{|x-\mu_j|^{2\al_j}}
{(x-z)(x-\mu_j)}f(x)dx=\\
{\pi\over\sin{2\pi\al_j}}(z-\mu_j)^{2\al_j-1}f(z)-
{\ep^{2\al_j}\over{2\al_j(z-\mu_j)}}f(\mu_j)-
{F_1(z)+O(\ep^{2\al_j+1})/(z-\mu_j)\over 2i\sin{2\pi\al_j}}.}
\]
The second is obtained by a similar analysis
with the cut of the root now stretching to the right hand side of $\mu_j$
and $x-\mu_j$ in the numerator replaced by $(x-\mu_j)e^{-i\pi}$.
We obtain:
\[
\eqalign{
\int_{\mu_j+\ep}^{d_j}\frac{|x-\mu_j|^{2\al_j}}
{(x-z)(x-\mu_j)}f(x)dx=\\
{-\pi e^{-2\pi i\al_j}\over\sin{2\pi\al_j}}(z-\mu_j)^{2\al_j-1}f(z)+
{\ep^{2\al_j}\over{2\al_j(z-\mu_j)}}f(\mu_j)+
{F_2(z)+O(\ep^{2\al_j+1})/(z-\mu_j)\over 2i\sin{2\pi\al_j}},}
\]
where $F_2(z)$ is regular as $z\to\mu_j$.

Now (\ref{pv}) can 
be written as follows (note, firstly, that $\ep^{2\al_j}$-contributions from the 
small circles around $\mu_j$ from the 2 integrals cancel each other,
and secondly, that $F_{1,2}(z)(z-\mu_j)=O(z-\mu_j)$):
\be\eqalign{
\int_{c_j}^{d_j}\frac{|x-\mu_j|^{2\al_j}}{x-z}f(x)dx-
\int_{c_j}^{d_j}\frac{|x-\mu_j|^{2\al_j}}{x-\mu_j}f(x)dx
=\\
e^{-i\pi\al_j}{\pi i\over \cos{\pi\al_j}}(z-\mu_j)^{2\al_j}f(\mu_j)
+O((z-\mu_j)^{2\al_j+1}),}\la{difference}
\ee
as $z\to\mu_j$, $\Im z>0$, $-1/2<\Re\al_j\le 0$, $\al_j\neq 0$. 
Thus, the difference of $Y_{j2}(z)$, $j=1,2$ and the principal value integrals
has a nonvanishing part at $\mu_j$ given by the first term in 
the r.h.s. of this equation. 
The difference vanishes for $\Re\al_j>0$
(as a similar analysis shows, this difference
 has only $(z-\mu_j)^{m}\ln (z-\mu_j)$-type singular terms with $m\ge 2k+1$ for
$\al_j=k+1/2$, $k=0,1,\dots$, $z\to\mu_j$,
and no irregular terms for other positive values of $\al_j$).
Finally, there is a constant difference
for $\al_j=0$.

We denote by $Y^{\rm vp}(\mu_j)$ 
the matrix $Y$ with the integrals of the second column
replaced by the principal value integrals. Namely,
\be\la{reg}
Y^{\rm vp}_{k2}(\mu_j)=
\lim_{z\to\mu_j}(Y_{k2}(z)-S_{k}(z)),
\ee
where the limit is taken along a
non-tangential to the real line path in $\bbc_+$, and 
\be\la{Sing}
S_k(z)=\cases{e^{-i\pi\al_j}{\pi i\over \cos{\pi\al_j}}(z-\mu_j)^{2\al_j}f_k(\mu_j),&
for $-1/2<\Re\al_j\le 0$, $\al_j\neq 0$\cr
\pi i f_k(\mu_j),& for $\al_j=0$\cr
0,& for $\Re\al_j>0$}
\ee
with 
\[
\eqalign{
f_1(\mu_j)=(2\pi i\ka_n)^{-1}p_n(\mu_j)
\prod_{s\neq j}|\mu_j-\mu_s|^{2\al_s}\exp(-\mu_j^2),\\
f_2(\mu_j)=-\ka_{n-1}p_{n-1}(\mu_j)\prod_{s\neq j}|
\mu_j-\mu_s|^{2\al_s}\exp(-\mu_j^2).}
\]

We then have by (\ref{D2}), (\ref{J1e}), (\ref{J2e}), and (\ref{RHM}),
\be\eqalign{
{d\over d\al}\ln D_n(\al_1,\dots,\al_m)=
-n(\ln\ka_{n-1})'_\al-\left(n+2\sum_{j=1}^m\al_j\right)(\ln\ka_n)'_\al-
2\left(\ka_{n-1}\over \ka_n\right)^2
\left(\ln{\ka_{n-1}\over\ka_n}\right)'_\al+\\
2\sum_{j=1}^m\al_j\left(\ka_n^{-1}(\ka_n Y_{11}(\mu_j))'_\al Y^{\rm vp}_{22}(\mu_j)-
\ka_{n-1}(\ka_{n-1}^{-1}Y_{21}(\mu_j))'_\al Y^{\rm vp}_{12}(\mu_j)\right)+
2\left[\ga'_{n,\al}-\beta_n\beta'_{n,\al}\right].}
\ee

This can be somewhat simplified by writing out the derivatives of 
$\ka_n Y_{11}(\mu_j)$,
$\ka_{n-1}^{-1}Y_{21}(\mu_j)$, and using the identity
$1=\det Y(\mu)=\det Y^{\rm vp} (\mu)= Y_{11}(\mu)Y^{\rm vp}_{22}(\mu)-
 Y_{21}(\mu)Y^{\rm vp}_{12}(\mu)$.

We obtain then, recalling that $\al=\al_\nu$, the following

\noindent
{\bf Proposition.} {\it Let $\al_j$, $\Re\al_j>-1/2$, $j=1,\dots,m$ be such that the
system of orthogonal polynomials $p_k(z)$ with finite nonzero leading coefficients 
satisfying (\ref{OP}) exists. Fix $n>1$. Let $p_n=\ka_n(x^n+\beta_n x^{n-1}+\gamma_n
x^{n-2}+\cdots)$, the matrix $Y$ be given by (\ref{RHM}), and $Y^{\rm vp}$ 
be defined in (\ref{reg}). Then   
\be\eqalign{
{d\over d\al_\nu}\ln D_n(\al_1,\dots,\al_m)=
-\left(n+2\sum_{j=1}^m\al_j\right)(\ln\ka_n\ka_{n-1})'_{\al_\nu}-
2\left(\ka_{n-1}\over \ka_n\right)^2
\left(\ln{\ka_{n-1}\over\ka_n}\right)'_{\al_\nu}+\\
2\sum_{j=1}^m\al_j\left(Y_{11}(\mu_j)'_{\al_\nu} Y^{\rm vp}_{22}(\mu_j)-
Y_{21}(\mu_j)'_{\al_\nu} Y^{\rm vp}_{12}(\mu_j)
+(\ln\ka_n\ka_{n-1})'_{\al_\nu}Y_{11}(\mu_j) Y^{\rm vp}_{22}(\mu_j)\right)+\\
2\left[\ga'_{n,\al_\nu}-\beta_n\beta'_{n,\al_\nu}\right],\qquad
\nu=1,\dots,m.}\la{id}
\ee
}

Note that $\ka^2_{n-1}=\lim_{z\to\infty}{iY_{21}(z)\over{2\pi
z^{n-1}}}$ and $\ga_n$, $\beta_n$ are expressed similarly.
Thus we can find the r.h.s. of (\ref{id}) asymptotically as
$n\to\infty$ provided the asymptotics of $Y(z)$ are available.
These are found in the next section.

\section{Asymptotic analysis of the Riemann-Hilbert problem}

In this section we are guided by \ci{KV,Dstrong} where the necessary
steepest-descent analysis of the problem (a)-(d) of Section 2 was
carried out. We are now faced only with a straightforward, although
cumbersome, calculation. We shall see that in order to
obtain asymptotics of the r.h.s. of (\ref{id}) to the needed accuracy,
we have to compute the first two terms in the
asymptotics of the coefficients $\ka_n$, $\beta_n$, $\ga_n$; and only
the main term in the asymptotics of $Y(\mu_j)$.
In this section we assume only the condition $\Re\al_j>-1/2$ for $\al_j$.

For the rest of the paper, we assume that 
\[
\lb_j\in(-1,1),\qquad j=1,\dots,m,\qquad \lb_j\neq\lb_k
\quad \mathrm{if}\quad j\neq k.
\]

\subsection{$U$, $T$, and $S$ transformations of the Riemann-Hilbert problem} 

As usual, we perform a series of transformations of the initial
problem for $Y(z)$. The first one $Y\to U$ is a scaling:
\be\la{U}
Y(z\sqrt{2n})=(2n)^{n\si_3/2}U(z),\qquad \si_3=\pmatrix{1&0\cr 0&-1}.
\ee  
The second one $U\to T$ is given by the formula
\be\la{T}
U(z)=(2n)^{\A\si_3/2}e^{nl\si_3/2}T(z)e^{n(g(z)-l/2)\si_3}(2n)^{-\A\si_3/2},
\ee
where
\be\la{gf}\eqalign{
\A=\sum_{j=1}^m\al_j,\qquad l=-1-2\ln 2,\\
g(z)=\int_{-1}^1\ln(z-s)\psi(s)ds,\qquad z\in\bbc\setminus(-\infty,1],\qquad
\psi(z)={2\over\pi}\sqrt{1-z^2}.}
\ee
Below we
always take the principal branch of the logarithm and roots.
The function $\psi(z)$
is the scaled asymptotic density of zeros of $p_n(z)$. (Note that $\int_{-1}^1\psi(x)dx=1$.)
The factor $e^{n g(z)}$ can therefore be regarded as a rough approximation for the polynomials.
The function $g(z)$ has the following useful properties:
\be\la{g}
\eqalign{
g_+(x)+g_-(x)-2x^2-l=0,\qquad \mathrm{for}\quad x\in(-1,1)\\
g_+(x)+g_-(x)-2x^2-l<0,\qquad \mathrm{for}\quad x\in\bbr\setminus[-1,1]\\
g_+(x)-g_-(x)=\cases{2\pi i, & for $x\le-1$\cr
2\pi i\int_x^1\psi(y)dy, & for $x\in [-1,1]$\cr
0, & for $x\ge 1$}}
\ee

From the Riemann-Hilbert problem for $Y(z)$ we obtain the following problem for $T(z)$:

\begin{enumerate}
    \item[(a)]
        $T(z)$ is  analytic for $z\in\bbc \setminus\bbr$.
    \item[(b)]  
The boundary values of $T(z)$ are related by the jump condition
\begin{equation}
\eqalign{   T_+(x) = T_-(x)
            \pmatrix{
                e^{-n(g_+(x)-g_-(x))} & \prod_{j=1}^m|x-\lb_j|^{2\al_j} \cr
                0 & e^{n(g_+(x)-g_-(x))}},
            \qquad\mbox{$x\in(-1,1)\setminus\cup_{j=1}^m\{\lb_j\}$.}\\
            T_+(x) = T_-(x)
            \pmatrix{
                1 & \prod_{j=1}^m|x-\lb_j|^{2\al_j}e^{n(g_+(x)+g_-(x)-2x^2-l)}\cr
                0 & 1},
            \qquad\mbox{$x\in\bbr\setminus[-1,1]$.}}
        \end{equation}
    \item[(c)]
$T(z)=I+O(1/z)$ as $z\to\infty$.
\item[(d)]
Behavior of $T(z)$ as $z\to\lb_j$ is the same as for $Y(\sqrt{2n}z)$.
\end{enumerate}
Note that this problem is normalized to $1$ at infinity, and 
the jump matrix on $(-\infty,-1)\cup (1,\infty)$ is exponentially close
to the identity (see (\ref{g})). We have to exclude small neighborhoods of the 
points $-1$ and $1$,
where $g_+(x)+g_-(x)-2x^2-l$ is close to zero, to have a uniform bound.

Now let $h(z)$ be the analytic continuation of 
\be
h(x)=g_+(x)-g_-(x)=2\pi e^{3i\pi/2}\int_1^x\psi(y)dy
\ee
to $\bbc\setminus((-\infty,-1]\cup[1,\infty))$.
Then a simple analysis shows that $\Re h(z)>0$ for $\Im z>0$,
and $\Re h(z)<0$ for $\Im z<0$ in some neighborhood of $(-1,1)$. We again exclude 
neighborhoods of the points $-1$ and $1$ for a uniform estimate.

It turns out it is possible to make use of this fact and split the contour
on $(-1,1)$ as shown in
Figure 2., transforming again the Riemann-Hilbert problem accordingly
(steepest descent method of Deift and Zhou). 
\begin{figure}
\centerline{\psfig{file=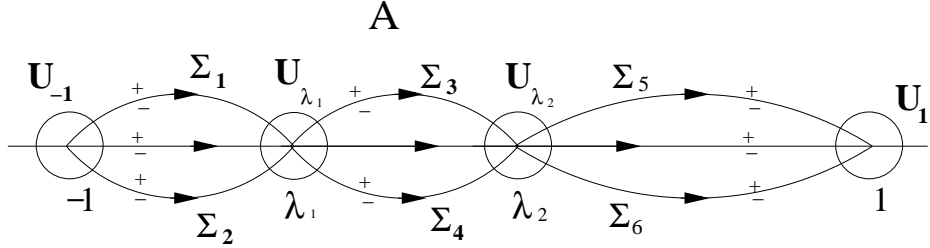,width=5.0in,angle=0}}
\vspace{0cm}
\caption{
Contour for the Riemann-Hilbert problem ($m=2$).}
\label{fig1}
\end{figure}
Then more elements of the jump matrix become exponentially small in
$n$, and hence asymptotically negligible, and what remains can be
reduced to matrices with constant (in $z$) elements. These jump
matrices are different in the $m+3$ neighborhoods $U_{\pm1}$, $U_{\lb_j}$
(discs around the points $\pm1$ and $\lb_j$ of some sufficiently small fixed 
radius $\delta$), and 
$U_\infty=\bbc\setminus(\overline{U_1\cup U_{-1}\cup_{j=1}^m U_{\lb_j}})$. 
Note that in $U_{\lb_j}$ we will have to add an additional cut roughly
perpendicular to the real axis (because of nonanalyticity of the absolute value).
Our task is then to construct approximate solutions
(parametrices) $P_{\pm1}(z)$, $P_{\lb_j}(z)$, $P_\infty (z)$ in the
corresponding neighborhoods
and match them on the boundaries $\partial U_{\pm 1}$, $\partial
U_{\lb_j}$. The solution in $U_\infty$ is given in terms of elementary
functions; in $U_{\pm1}$, in terms of Airy functions \ci{Dstrong};
and in $U_{\lb_j}$, in terms of Bessel functions \ci{KV,V}. The
``matching'' is performed using the known asymptotics of these
functions at large arguments
and one more ``$R$'' transformation of the Riemann-Hilbert
problem. This allows us to compute any number of terms in the
asymptotics of $Y(z)$.

We start by splitting the contour $(-1,1)$ into lenses (Figure 2).
Define an analytic continuation of $\prod_{j=1}^m|x-\mu_j|^{2\al_j}$ as follows:
\be
\om(z)=\cases{\prod_{j=1}^m(\lb_j-z)^{2\al_j},& for $\Re z<\lb_1$,\cr
\prod_{j=1}^\nu(z-\lb_j)^{2\al_j}\prod_{j=\nu+1}^m(\lb_j-z)^{2\al_j},& for 
$\lb_\nu<\Re z<\lb_{\nu+1}$,  $\nu=1,\dots,m-1$\cr
\prod_{j=1}^m(z-\lb_j)^{2\al_j},& for $\Re z>\lb_m$.}
\ee

Splitting of the contour is possible because of the following factorization
property of the jump matrix on $(-1,1)$: 
\be
\pmatrix{e^{-nh(x)}& \om(x)\cr 0& e^{nh(x)}}=
\pmatrix{1&0\cr \om(x)^{-1}e^{nh(x)}& 1}
\pmatrix{0 & \om(x)\cr -\om(x)^{-1} & 0}
\pmatrix{1&0\cr \om(x)^{-1}e^{-nh(x)}& 1}.
\ee

Define a new transformation of our matrix-valued function as follows:
\be\la{defS}
S(z)=
\cases{T(z),& for $z$ outside the lenses,\cr
T(z)\pmatrix{1 & 0\cr -\om(z)^{-1}e^{-nh(z)}& 1}, & 
for $z$ in the upper part of the lenses,\cr
 T(z)\pmatrix{1 & 0\cr \om(z)^{-1}e^{nh(z)}& 1},&
for $z$ in the lower part of the lenses.
}
\ee
(Note that we had to contract the lenses at $\lb_j$ because 
we want $S(z)$ to be analytic
in each part of the lenses.) 

Then the Riemann-Hilbert problem for $S(z)$ is the following:

\begin{enumerate}
    \item[(a)]
        $S(z)$ is  analytic for $z\in\bbc \setminus\Sigma$, where 
$\Si=\bbr\cup \cup_{j=1}^{2m+2}\Si_j$.
    \item[(b)]  
The boundary values of $S(z)$ are related by the jump condition
\begin{equation}
\eqalign{   S_+(x) = S_-(x)
            \pmatrix{
                  1 & 0\cr
                  \om(x)^{-1}e^{\mp n h(x)} & 1},
            \qquad\mbox{$x\in\cup_{j=1}^{2m+2}\Sigma_j$},\\
\mbox{where the plus sign in the exponent is on 
$\Sigma_{2j}$, and minus, on $\Sigma_{2j-1}$, $j=1,2,\dots,m+1$,}\\
            S_+(x) = S_-(x)
              \pmatrix{
                  0 & \om(x)\cr
                  -\om(x)^{-1} & 0},
            \qquad\mbox{$x\in(-1,1)\setminus\cup_{j=1}^m\{\lb_j\}$.}\\
            S_+(x) = S_-(x)
            \pmatrix{
                1 & \om(x)e^{n(g_+(x)+g_-(x)-2x^2-l)}\cr
                0 & 1},
            \qquad\mbox{$x\in\bbr\setminus[-1,1]$.}}
        \end{equation}
    \item[(c)]
$S(z)=I+O(1/z)$ as $z\to\infty$.
\item[(d)]
For $\Re\al_j\le 0$, the matrix function $S(z)$ has the following behavior as
$z\to\lb_j$:
        \begin{equation}\la{Sloc1}
            S(z)=
            O\pmatrix{
                1 & |z-\lb_j|^{2\al_j} \cr
                1 & |z-\lb_j|^{2\al_j}
            }, \qquad\mbox{as $z\to \lb_j, z\in\mathbb{C}\setminus\Sigma$.}
        \end{equation}
        For $\Re\al_j>0$, the matrix function $S(z)$ has the
        following behavior as $z\to\lb_j$:
        \begin{equation}\la{Sloc2}
            S(z)=\left\{\begin{array}{cl}
                O\pmatrix{
                    1 & 1 \cr
                    1 & 1
         },& \mbox{as $z\to\lb_j$ from outside the lenses,} \\
                O\pmatrix{
                    |z-\lb_j|^{-2\al_j} & 1 \cr
                    |z-\lb_j|^{-2\al_j} & 1
                }, & \mbox{as $z\to \lb_j$ from inside the lenses.}
            \end{array}\right.
        \end{equation}
\end{enumerate}

Recalling the remarks above, we see that,
outside the neighborhoods $U_{\lb_j}$, $U_{\pm1}$,
the jump matrix on $\Sigma_j$, $j=1,\dots,2m+2$ is uniformly exponentially close to 
the identity.

Let us now start constructing parametrices which give (in their respective regions)
the leading contribution to the asymptotics.

\subsection{Parametrix in $U_\infty$}

First, we expect the following problem for the parametrix in $U_\infty$:

\begin{enumerate}
    \item[(a)]
        $P_\infty(z)$ is  analytic for $z\in\bbc \setminus[-1,1]$,
    \item[(b)]
with the jump condition on $(-1,1)$
\be
P_{\infty,+}(x) = P_{\infty,-}(x)
            \pmatrix{0&\om(x)\cr
              -\om(x)^{-1}&0},
\qquad\mbox{$x \in (-1,1)\setminus\cup_{j=1}^m\{\lb_j\}$},
\ee
\item[(c)]
and the following behavior at infinity
\be
P_\infty(z) = I+ O \left( \frac{1}{z} \right), 
     \qquad \mbox{as $z\to\infty$.}
\ee
\end{enumerate}
A solution $P_\infty(z)$ can be found in the same way as in \ci{KVA}.
\be
P_\infty(z)={1\over 2}({\cal D}_\infty)^{\si_3}
\pmatrix{a+a^{-1}&-i(a-a^{-1})\cr i(a-a^{-1})& a+a^{-1}}
{\cal D}(z)^{-\si_3},
\qquad
a(z)=\left({z-1\over z+1}\right)^{1/4},\la{N}
\ee
where the cut of the root is the interval $(-1,1)$. 
(Note that $\det P_\infty(z)=1$.)
Here
\be
{\cal D}(z)=\exp\left[{\sqrt{z^2-1}\over 2\pi}\int_{-1}^1
{\ln \om(\xi)\over \sqrt{1-\xi^2}}
{d\xi \over z-\xi}\right],\qquad {\cal D}_\infty=\lim_{z\to\infty}
{\cal D}(z).\la{Sf}
\ee

The Szeg\H o function ${\cal D}(z)$ is
analytic outside the interval $[-1,1]$
with the boundary values satisfying ${\cal D}_+(x){\cal D}_-(x)=\om(x)$,
$x\in (-1,1)\setminus\cup_{j=1}^m\{\lb_j\}$. 
We can calculate 
the integral in (\ref{Sf}) directly as follows. First, replace
$\om(x)$ by $\om(\nu x)$; second, take the derivative of the integral
w.r.t. $\nu$. Set then 
$\xi=\cos\theta$, and integrate in the complex plane of
$e^{i\theta}$. After that integrate the result w.r.t. $\nu$ over
$(0,1)$, and finally, use the value of the integral at $\nu=0$. 
This procedure is actually a simple ``scalar'' analogue of the whole
present work.
For $\lb_j\in (-1,1)$, we obtain
\be\la{Dom}
{\cal D}(z)=(z+\sqrt{z^2-1})^{-\A}\prod_{j=1}^m(z-\lb_j)^{\al_j},\qquad 
\A=\sum_{j=1}^m\al_j. 
\ee
Therefore,
\be
{\cal D}_\infty=2^{-\A}.
\ee

%%%%%%%%%%%%% \lb_j %%%%%%%%%%%%%%%%%%%%%%%%%%%%%%%%%%%%%%%%

\subsection{Parametrices at $z=\lb_j$}

Let us now construct the parametrices in $U_{\lb_j}$, $j=1,\dots,m$. 
The construction is the same for every $j$.
We look for an analytic matrix-valued function $P_{\lb_j}(z)$ in a 
neighborhood of $U_{\lb_j}$
which satisfies the same jump conditions as $S(z)$ on
$\Sigma\cap U_{\lb_j}$, has the same behavior as $z\to\lb_j$, and 
(instead of a condition at infinity)
satisfies the matching condition
\be\la{match}
P_{\lb_j}(z)P_\infty^{-1}(z)=I+O(1/n)
\ee
uniformly on the boundary $\partial U_{\lb_j}$ as $n\to\infty$.

Using the analytic continuation of $\psi(y)$, define:
\be
\phi(z)=\cases{h(z)/2=e^{3i\pi/2}\pi\int_1^z\psi(y)dy,& for $\Im z>0$,\cr
e^{-i\pi}h(z)/2=e^{i\pi/2}\pi\int_1^z\psi(y)dy,& for $\Im z<0$},
\ee
It is easy to verify that the function $e^{\phi(z)}$ is analytic 
outside $[-1,1]$. Note that $h(z)$ was
defined analytic outside $\bbr\setminus(-1,1)$.
Set furthermore,
\be\la{f}
\hat f(z)=\pi\int_{\lb_j}^z\psi(y)dy.
\ee
Let us now choose the exact form of the cuts $\Sigma$ in $U_{\lb_j}$ so that their images
under the mapping $\zeta=n\hat f(z)$ are direct lines (Figure 3).
Note that $\zeta(z)=n\hat f(z)$ is analytic and one-to-one in the neighborhood 
of $U_{\lb_j}$, and it takes the real axis to the real axis.
We have:
\be\la{zlb}
\zeta=n \hat f(z)=n2\sqrt{1-\lb_j^2}(z-\lb_j)(1+O(z-\lb_j)),\qquad z\to\lb_j.
\ee

Set
\be\la{W}
W_j(z)=\prod_{i=1}^{j-1}(z-\lb_i)^{\al_i}\prod_{i=j+1}^{m}(\lb_i-z)^{\al_i}\times
\cases{(z-\lb_j)^{\al_j},& if $\pi/2<|\arg \hat f(z)|<\pi$\cr
(\lb_j-z)^{\al_j},& if $0<|\arg \hat f(z)|<\pi/2$}.
\ee
This function has the following jumps on $\Ga_3$, $\Ga_7$:
\be
W_{j+}(z)=W_{j-}(z)e^{i\pi\al_j},\qquad \ze\in \Ga_3\cup\Ga_7.
\ee
Moreover, the functions $W_j(z)$ and $\om(z)$ are related in $U_{\lb_j}$ as follows:
\be
W^2_j(z)=\om(z)e^{-2\pi i\al_j}
\ee
in the region $\Re \ze>0$, $\Im z>0 \cap \Im\ze>0$, and in the region
$\Re \ze<0$, $\Im z<0 \cap \Im\ze<0$;
whereas
\be
W^2_j(z)=\om(z)e^{2\pi i\al_j}
\ee
in the region $\Re \ze<0$, $\Im z>0 \cap \Im \ze>0$, and in the region
$\Re \ze>0$, $\Im z<0 \cap \Im \ze<0$.
Note that by definition of $\om(z)$, $\arg\om(z)=0$ on $\Im z=0$, and $\om(z)$ is 
continuous through this line. 
Hence the function $W_j(z)$ also has jumps on $\Ga_1$, $\Ga_5$ given by the formulae:
\be
W_{j+}(z)=W_{j-}(z)\times \cases{e^{-2\pi i\al_j},& for $\ze\in\Ga_1$\cr
                           e^{2\pi i\al_j},& for $\ze\in\Ga_5$}.
\ee

We look for $P_{\lb_j}(z)$ in the form:
\be\la{Plb0}
P_{\lb_j}(z)=E_n(z)P^{(1)}(z)W_{j}(z)^{-\si_3}e^{-n\phi(z)\si_3},
\ee
where $E_n(z)$ is analytic and invertible in the neighborhood of $U_{\lb_j}$,
and therefore does not affect the jump and analyticity conditions. 
It is needed and so chosen that the matching condition be satisfied.

It is easy to verify that $P^{(1)}(z)$ satisfies jump conditions with {\it constant}
jump matrices. Because of $W_{j}(z)$, $P^{(1)}(z)$ has also an
{\it additional} jump condition along the line $\Re \hat f(z)=0$ in $U_{\lb_j}$.
Set
\be
P^{(1)}(z)=\Psi_{\al_j}(\zeta)=\Psi_{\al_j}(n \hat f(z)),
\ee
where the cuts for $\Psi_{\al_j}(\zeta)$ are shown in Figure 3, 
and the Riemann Hilbert problem 
for it is as follows:

\begin{figure}
\centerline{\psfig{file=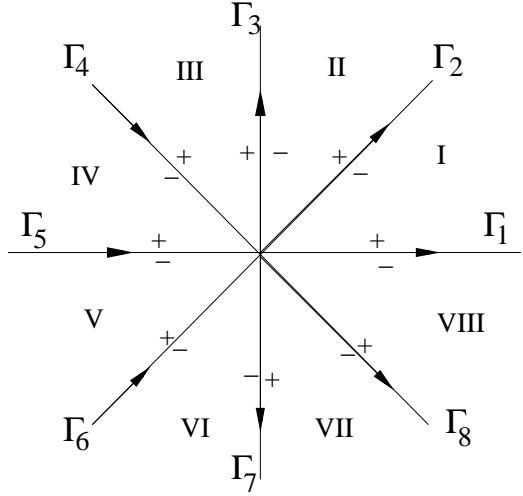,width=2.7in,angle=0}}
\vspace{0cm}
\caption{
The auxiliary contour for the parametrix at $\lb_j$.}
\label{fig2}
\end{figure}

\begin{enumerate}
\item[(a)]
    $\Psi_{\al_j}$ is analytic for $\ze\in\bbc\setminus\cup_{j=1}^8\Ga_j$.
\item[(b)]
    $\Psi_{\al_j}$ satisfies the following jump conditions:
    \begin{eqnarray} \label{jumpPsi}
        \Psi_{\al_j,+}(\zeta)
        &=&
            \Psi_{\al_j,-}(\zeta)
            \pmatrix{
                0 & 1 \cr
                -1 & 0
            },
            \qquad \mbox{for $\zeta \in \Gamma_1\cup\Gamma_5$,} \\
        \Psi_{\al_j,+}(\zeta)
        &=& 
            \Psi_{\al_j,-}(\zeta)
            \pmatrix{
                1 & 0 \cr
                e^{-2\pi i \al_j}  & 1
            },
            \qquad \mbox{for $\zeta \in \Gamma_2\cup\Gamma_6$,} \\
        \Psi_{\al_j,+}(\zeta)
        &=& 
            \Psi_{\al_j,-}(\zeta)
            e^{\pi i \al_j\sigma_3},
            \qquad \mbox{for $\zeta \in \Gamma_3\cup\Gamma_7$,} \\
        \Psi_{\al_j,+}(\zeta)
        &=& 
            \Psi_{\al_j,-}(\zeta)
            \pmatrix{
                1 & 0 \cr
                e^{2\pi i \al_j}  & 1
            },
            \qquad \mbox{for $\zeta \in \Gamma_4\cup\Gamma_8$.}
    \end{eqnarray}
\item[(c)]
    For $\Re\al_j\le 0$ the matrix function $\Psi_{\al_j}(\zeta)$ has the 
following behavior as $\zeta\to 0$:
    \begin{equation}\la{Psiloc1}
        \Psi_{\al_j}(\zeta)=
        O\pmatrix{
            |\zeta|^{\al_j} & |\zeta|^{\al_j} \cr
            |\zeta|^{\al_j} & |\zeta|^{\al_j}
        },
        \qquad \mbox{$\zeta\to 0$.}
    \end{equation}
    For $\Re\al_j>0$ the matrix function $\Psi_{\al_j}(\zeta)$ has the following 
behavior as $\zeta\to 0$:
    \begin{equation}\la{Psiloc2}
        \Psi_{\al_j}(\zeta)=
        \left\{\begin{array}{cl}
            O\pmatrix{
                |\zeta|^{\al_j} & |\zeta|^{-\al_j} \cr
                |\zeta|^{\al_j} & |\zeta|^{-\al_j}
            },
            & \mbox{as $z\to\lb_j$ with $\zeta\in$ II, III, VI, VII,} \\
            O\pmatrix{
                |\zeta|^{-\al_j} & |\zeta|^{-\al_j}\cr
                |\zeta|^{-\al_j} & |\zeta|^{-\al_j}
            },
            & \mbox{as $z\to \lb_j$ with $\zeta\in$ I, IV, V, VIII.}
        \end{array}\right.
    \end{equation}
\end{enumerate}

The solution of this Riemann-Hilbert problem was
constructed in \ci{V} in terms of Bessel functions. 
For our purposes, we need its explicit form only in the region $II$ (see Figure 3).
There, we have (see \ci{V}):
 
\be\la{Psi}
\Psi_{\al_j}(\ze)=\sqrt{\pi\ze}\pmatrix{
I_{\al_j+1/2}(\ze e^{-i\pi/2})e^{-i\pi\al_j/2}& 
-{1\over\pi}K_{\al_j+1/2}(\ze e^{-i\pi/2})e^{i\pi\al_j/2}\cr
-i I_{\al_j-1/2}(\ze e^{-i\pi/2})e^{-i\pi\al_j/2}&
-{i\over\pi}K_{\al_j-1/2}(\ze e^{-i\pi/2})e^{i\pi\al_j/2}},\quad \zeta\in II,
\ee
where $I_\beta(x)$ and $K_\beta(x)$ are modified Bessel functions (see
\ci{Abr}).
The solution in other regions can be reproduced by applying the jump conditions to
(\ref{Psi}).

Taking also $E_n(z)$ from \ci{KV,V}, and substituting all into (\ref{Plb0}),
we obtain for $z\in z(II)$, where $z(II)$ is the image of $II$ under the 
mapping $\ze\rightarrow z$, 
\be\la{Plb}
P_{\lb_j}(z)=P_\infty(z)W_{j}(z)^{\si_3}e^{(n\phi_+(\lb_j)+\al_j\pi i/2)\si_3}
e^{-i\pi\si_3/4}{1\over\sqrt{2}}\pmatrix{1& i\cr i& 1}
\Psi_{\al_j} (n\hat f(z))W_{j}(z)^{-\si_3}e^{-n\phi(z)\si_3}.
\ee

The argument of Bessel functions is uniformly large on $\partial U_{\lb_j}$.  
Substituting asymptotics of Bessel functions for large arguments 
\be
\Psi_{\al_j}(\ze)={1\over\sqrt{2}}\pmatrix{1 & -i \cr -i & 1}
\left[ I+{i\over 4\ze}\pmatrix{-2\al_j^2& -2i\al_j\cr -2i\al_j& 2\al_j^2}+
O(\ze^{-2})\right]e^{(\pi/4-\al_j\pi/2-\ze)i\si_3}, \ze\in II,
\ee
(and similar ones for other regions (see \ci{KV,V}))
into (\ref{Plb}) (and its analogues for other regions), 
we can verify that the matching condition is satisfied.
For that we need only the main asymptotic term plus the error term. The fit is ensured by 
the choice of $E_n(z)$. The exponential factor in the asymptotics of Bessel functions 
cancels with $e^{-n\phi(z)\si_3}$ in (\ref{Plb}) leaving a constant in $z$ factor.
(Note that $\hat f(z)=i\phi(z)-i\phi_+(\lb_j)$.)
Moreover, considering further terms, we can extend
(\ref{match}) into full asymptotic series in inverse powers of $n$.
For our calculations we need to know explicitly the first correction term:
\be\la{alb}\eqalign{
P_{\lb_j}(z)P_\infty^{-1}(z)=I+\De_1(z)+O(1/n^2),\\
\De_1(z)=-P_\infty(z)W_{j}(z)^{\si_3}e^{(n\phi_+(\lb_j)+\al_j\pi i/2-\pi i/4)\si_3}
{i\over 2\ze}\pmatrix{\al_j^2& i\al_j\cr i\al_j& -\al_j^2}\times\\
W_{j}(z)^{-\si_3}e^{-(n\phi_+(\lb_j)+\al_j\pi i/2-\pi i/4)\si_3}P_\infty^{-1}(z),\\
\qquad z\in\partial z(II),}
\ee
where $\partial z(II)$ is the 
part of $\partial U_{\lb_j}$ whose $\ze$-image is in $II$.
As the calculation for the other regions shows,
this expression for $\De_1(z)$ extends by analytic continuation to the whole boundary
$\partial U_{\lb_j}$ (cf. \ci{KV,V}). Moreover, it gives a meromorphic function in 
a neighborhood of $U_{\lb_j}$ with a simple pole at $z=\lb_j$.
The error term $O(1/n^2)$ in (\ref{alb}) is uniform on $\partial U_{\lb_j}$.

Note that the absence of jumps for $\det\Psi(\ze)$ and the fact that $\Re\al_j>-1/2$
implies that the only possible singularity of $\det\Psi(\ze)$ (at $\ze=0$) is removable.
Thus, $\det\Psi(\ze)$ is analytic.
Moreover, the asymptotics of Bessel functions give that $\det\Psi(\ze)\to 1$ as $\ze\to\infty$,
which implies that $\det\Psi(\ze)\equiv 1$. Using this, we easily deduce from (\ref{Plb}) 
that also $\det P_{\lb_j}(z)\equiv 1$.

Note that it follows from the asymptotics of Bessel functions 
and the fact that
\be\la{DW}
D(z)W_{j}(z)^{-1}=(z+\sqrt{z^2-1})^{-\A}e^{\pm i\pi\sum_{k=j'}^m\al_k}
\ee
(where $j'=j$ or $j'=j+1$ and the sign is different in different 
quadrants of the $\ze$-plane),
that all the terms (including the error term)
are uniform for $\al_k$ in a bounded set of the half-plane $\Re\al_k>-1/2$.
This observation (and similar ones for
$\partial U_{\pm1}$) will be very important below.

%%%%% 1 %%%%%%%%%%

\subsection{Parametrices at $z=\pm1$}

Now let us construct parametrices in the remaining regions $U_{\pm1}$.
These are obtained by a slight generalization
of the results of \cite{Dstrong} (which can be viewed as the case $\om(z)=1$).
We are looking for an analytic matrix-valued function in $U_1$ which has
the same jump
relation as $S(z)$ there and satisfies the matching 
condition on the boundary:
\be\la{match2}
P_1(z)P_\infty^{-1}(z)=I+O(1/n).
\ee

The solution is:
\be
P_1=E(z)Q(\ze)e^{-n\phi(z)\si_3}\om(z)^{-\si_3/2},\qquad
E(z)=P_\infty(z)\om(z)^{\si_3/2}e^{i\pi\si_3/4}\sqrt{\pi}
\pmatrix{1&-1\cr 1&1}\ze^{\si_3/4}e^{-\pi i/12},
\ee
and $Q(\ze)$ is given by the expression (7.9) of \cite{Dstrong}
in terms of Airy functions
(in the notation of \cite{Dstrong} $Q(\ze)=\Psi^\si(\zeta)$).
In these formulas
\be
\ze(z)=\left( {3\over 2}n e^{-i\pi}\phi(z)\right)^{2/3}.
\ee
As is easy to verify, so defined $\ze(z)$ is an analytic function 
in a neighborhood of $z=1$ (the cut of the square root is the
interval $(-1,1)$), and
\be
\ze(z)=2 n^{2/3} (z-1) \left(1+{1\over 10} (z-1)+ O((z-1)^2)\right).
\ee

The argument of Airy function on $\partial U_1$ is uniformly large, 
so we can expand it into the 
asymptotic series and proceed the same way as for $\partial U_{\lb_j}$. As a result
we have the matching condition (\ref{match2}) extended to the full asymptotic 
expansion in inverse powers of $n$.
We shall need below only the first 2 terms:
\be\la{a1}\eqalign{
P_1(z)P_\infty^{-1}(z)=I+\De_1(z)+O(1/n^2),\\
\De_1(z)=P_\infty(z)\om(z)^{\si_3/2}e^{\pi i\si_3/4}
{1\over 12}\pmatrix{1/6 & 1\cr -1& -1/6}
e^{-\pi i\si_3/4}\om(z)^{-\si_3/2}P_\infty^{-1}(z)
{3\over 2}\ze^{-3/2},\\
z\in \partial U_1.}
\ee

The function $\De_1(z)$ is meromorphic in the neighborhood of $U_1$ with 
a second order pole at $z=1$.

%%%%%%%% -1 %%%%%%%%%%%%

The argument for the parametrix in $U_{-1}$ is similar. 
We just mention the solution:
\be\la{P-1}\eqalign{
P_{-1}=E(z)\si_3 Q(e^{-i\pi}\zeta)\si_3 
e^{-n\tilde\phi(z)\si_3}\om(z)^{-\si_3/2},\\
E(z)=P_\infty(z)\om(z)^{\si_3/2}e^{i\pi\si_3/4}\sqrt{\pi}
\pmatrix{1&1\cr -1&1}(e^{-i\pi}\zeta)^{\si_3/4}e^{-\pi i/12},}
\ee
Here
\be
\zeta(z)=e^{-i\pi}\left( {3\over 2}n \tilde\phi(z)\right)^{2/3},\qquad
\tilde\phi(z)=
\cases{h(z)/2-i\pi=e^{3i\pi/2}\pi\int_{-1}^z\psi(y)dy,& for $\Im z>0$,\cr
e^{i\pi}h(z)/2+i\pi=
e^{5i\pi/2}\pi\int_{-1}^z\psi(y)dy,& for $\Im z<0$}.
\ee
For $z\to -1$:
\be
\ze(z)=2 n^{2/3} (1+z) \left(1-{1\over 10} (1+z)+ O((1+z)^2)\right).
\ee

The first 2 terms in the matching condition are as follows:
\be\la{am1}\eqalign{
P_{-1}(z)P_\infty^{-1}(z)=I+\De_1(z)+O(1/n^2),\\
\De_1(z)=P_\infty(z)\om(z)^{\si_3/2}e^{\pi i\si_3/4}
{1\over 12}\pmatrix{1/6 & -1\cr 1& -1/6}
e^{-\pi i\si_3/4}\om(z)^{-\si_3/2}P_\infty^{-1}(z)
{3\over 2}(e^{-i\pi}\ze)^{-3/2},\\
z\in\partial U_{-1}.}
\ee

As in (\ref{alb}), the error term in (\ref{a1}) (resp., (\ref{am1})) is uniform for 
all $z\in \partial U_{1}$ (resp., $z\in \partial U_{-1}$) and $\al_j$'s 
in a bounded set.

%%%%%%%%%%% R %%%%%%%%%%%%%

\subsection{Final transformation of the problem}

Now the construction of the parametrices is complete, and we are ready
for the final transformation of the Riemann-Hilbert
problem. Let
\be
R(z)=\cases{S(z)P_\infty^{-1}(z),& 
$z\in U_\infty\setminus\Si$,\cr
S(z)P_{\lb_j}^{-1}(z),& 
$z\in U_{\lb_j}\setminus\Si$, $j=1,\dots,m,$\cr
S(z)P_1^{-1}(z),&
$z\in U_1\setminus\Si$,\cr
S(z)P_{-1}^{-1}(z),&
$z\in U_{-1}\setminus\Si$.}\la{Rb}
\ee
It is easily seen that this function has jumps only on 
$\partial U_{\pm1}$, $\partial U_{\lb_j}$, and parts of $\Si_j$, $\bbr\setminus [-1,1]$
lying outside the neighborhoods $U_{\pm 1}$, $U_{\lb_j}$ (we
denote these parts without the end-points $\Si^\mathrm{out}$). 
The contour is shown in Figure 4 (for $m=2$). Outside of it, $R(z)$ is analytic:
Indeed, as follows from (\ref{Sloc1}), (\ref{Sloc2}), (\ref{Psiloc1}), 
(\ref{Psiloc2}), 
and the fact that $\det P_{\lb_j}=1$, the function
$S(z)P_{\lb_j}^{-1}(z)$, for example, has at most a singularity of order less than $1$
at $\lb_j$, which implies, due to the absence of jumps, that 
$S(z)P_{\lb_j}^{-1}(z)$ is analytic in $U_{\lb_j}$. 
The argument for other regions is similar.

Note that $R(z)=I+O(1/z)$ as $z\to\infty$. 

\begin{figure}
\centerline{\psfig{file=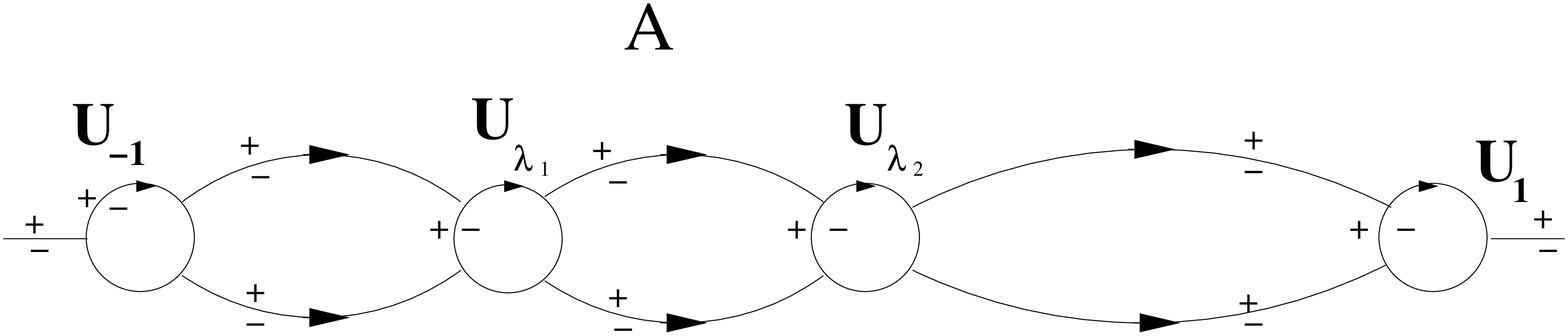,width=5.0in,angle=0}}
\vspace{0cm}
\caption{ 
Contour for the $R$-Riemann-Hilbert problem ($m=2$).}
\label{fig4}
\end{figure}

The jumps are as follows:
\be\eqalign{
R_+(x)=R_-(x)P_{\infty}(x)\pmatrix{1&\om(x)e^{n(g_+(x)+g_-(x)-2x^2-l)}
\cr 0&1}P_{\infty}(x)^{-1},\qquad x\in\bbr\setminus [-1-\delta,1+\delta],\\
R_+(x)=R_-(x)P_{\infty}(x)\pmatrix{1&0\cr \om(x)^{-1}
e^{\mp n h(x)}&1}P_{\infty}(x)^{-1},\qquad x\in\Si_k^\mathrm{out},
\qquad k=1,\dots,2m+2,\\
\mbox{where the plus sign in the exponent is on 
$\Sigma_{2j}^\mathrm{out}$, and minus, on $\Sigma_{2j-1}^\mathrm{out}$, $j=1,\dots,m+1$,}\\
R_+(x)=R_-(x)P_{\lb_j}(x)P_\infty (x)^{-1},\qquad x\in \partial U_{\lb_j}\setminus
\mbox{\{ intersection points\}},
\qquad j=1,\dots,m,\\
R_+(x)=R_-(x)P_{\pm 1}(x)P_\infty (x)^{-1},\qquad x\in \partial U_{\pm 1}\setminus
\mbox{\{ intersection points\}}.}\la{Rj}
\ee
Here $\delta$ is the radius of $U_1$, $U_{-1}$.

The jump matrix on $\Si^\mathrm{out}$ can be 
estimated uniformly in $\al$ as $I+O(\exp(-\ep n|x|))$, where 
$\ep$ is a positive constant.
The jump matrices on $\partial U_{\lb_j,\pm1}$ admit a uniform expansion in 
inverse powers of $n$ (the first 2 terms of which are 
given by (\ref{alb}), (\ref{a1}), and (\ref{am1})):
\be\la{Deas}
I+\De_1(z)+\De_2(z)+\dots+\De_{k}(z)+O(n^{-k-1}).
\ee
Every $\De_j$ is of order $1/n^j$.
(The above expressions give us an
explicit form of $\De_1(z)$ in each of the neighborhoods.)

We look for $R(z)$ asymptotically in the 
form $R(z)\sim R_0(z)+R_1(z)+R_2(z)+\cdots$,
where $R_j(z)$, $j>0$,  is of the same order as $\De_j$.
It can be shown following Theorems 7.8--7.10 of \ci{Dstrong} that
for any $k\ge1$
\be\la{Ras}
R(z)=\sum_{j=0}^{k}R_j(z)+O(n^{-k-1}),\qquad R_0=I,\la{R}
\ee
uniformly for all $z$ and for $\al_j$ in a bounded set 
of the half-plane $\Re\al_j>-1/2$, $j=1,\dots,m$. The set can extend up to 
the boundary $\Re\al_j=-1/2$.

Moreover, we can substitute this asymptotic expansion
into (\ref{Rj}) and, collecting the terms of the same order, obtain:
\be\eqalign{
R_{0+}(x)+R_{1+}(x)+\cdots\sim
(R_{0-}(x)+R_{1-}(x)+\cdots)(I+\De_1(x)+\cdots),
\qquad x\in \partial U_{\lb_j,\pm1}.\\
R_{0+}(x)=R_{0-}(x)\quad\Rightarrow\quad R_0(z)=I,\\
R_{1+}(x)-R_{1-}(x)=\De_1(x),\\
R_{2+}(x)-R_{2-}(x)=R_{1-}(x)\De_1(x)+\De_2(x),\\
R_{k+}(x)-R_{k-}(x)=\sum_{j=1}^k R_{k-j,-}(x)\De_j(x),\qquad k=1,2,\dots}
\ee
The main term in the asymptotics of the polynomials 
is given therefore by
the parametrices at the appropriate points $z$.
The expressions for $R_k(z)$ follow from the Sokhotsky-Plemelj
formulas:
\be
R_1(z)={1\over 2\pi i}\int_{\partial U}
{\De_1(x)dx\over x-z},\qquad
R_2(z)={1\over 2\pi i}\int_{\partial U}
{R_{1-}(x)\De_1(x)+\De_2(x)\over x-z}dx,\quad\dots\la{plem}
\ee
$\partial U=\partial U_1\cup\partial U_{-1}\cup_{j=1}^m\partial U_{\lb_j}$.
Note that the contours are traversed in the negative direction.

For reader's convenience, we present a variant of the proof of (\ref{Ras}--\ref{plem}),
a combination of the arguments from \ci{Dstrong} and \ci{KVA}.
First, we need a bound on $R(z)$.
Let
\be
\De\equiv J-I,
\ee
where $J$ is the jump matrix for $R$ on $\Si_R\equiv\Si^\mathrm{out}\cup\partial U$ 
(see (\ref{Rj})).
The jump condition and the behaviour of
$R(z)$ at infinity imply that
\be\la{R0}
R(z)=I+C(R_-\De),\qquad z\in\bbc\setminus\Si_R,
\ee
where
\[
C(f)={1\over 2\pi i}\int_{\Si_R} f(s){ds\over s-z}
\]
is the Cauchy operator on $\Si_R$.
Hence,
\be\la{R-}
R_-(s)=I+C_-(R_-\De),
\ee
where $C_-(f)=\lim_{z\to s}C(f)$ as $z$ approaches a point $s\in\Si_R\setminus
\mbox{\{intersection points\}}$
from the $-$ side of $\Si_R$. It is known that $C_-$ is a bounded operator
from $L^2(\Si_R)$ to $L^2(\Si_R)$. Now defining the operator
\[
C_{\De}(f)=C_-(f\De),
\]
we represent (\ref{R-}) in the form
\be\la{RDe}
(I-C_{\De})(R_--I)=C_\De(I).
\ee
Since by the estimates above
$\De(s)=O(1/n)$ and $\De(s)=O(\exp(-\ep n|s|))$ (on $\Si^{\mathrm{out}}$)
for $n\to\infty$ uniformly in $\al_j$'s (in a bounded set), and $s\in\Si_R$,
the operator norm of $C_\De$ acting on $L^2(\Si_R)$,
$||C_\De||=O(1/n)$, and $I-C_{\De}$ is invertible by 
a Neuman series for $n$ sufficiently large. Moreover, $||C_\De(I)||_{L^2(\Si_R)}=O(1/n)$.
Thus (\ref{RDe}) gives
\be
R_-(s)=I+(I-C_{\De})^{-1}(C_\De(1)),
\ee
where
\be
||R_-(s)-I||_{L^2(\Si_R)}=O(1/n).
\ee
Hence, by (\ref{R0}),
\be\la{R2}
R(z)=I+C[\De+(I-C_{\De})^{-1}(C_\De(I))\De],
\ee
and a matrix norm for some $\ep_1>0$ and ${\rm{dist}}(z,\Si_R)\ge\ep_1$, 
\be
|R(z)-I|\le |C(\De)|+|C((R_--I)\De)|\le {c_1\over n}+
c_2 ||R_-(s)-I||_{L^2(\Si_R)}||\De||_{L^2(\Si_R)}\le {c_3\over n}
\ee
uniformly in $\al$ and $z$
for some $c_1,c_2,c_3>0$, and $n$ larger then some $n_0$.

To obtain a uniform estimate for all $z\in\bbc\setminus\Si_R$, we (for a $z$
with ${\rm{dist}}(z,\Si_R)<\ep_1$) deform the contour as shown in Figure  5.
Here $\ti\Si_R$ is the same as $\Si_R$ with the dotted part replaced by the
semicircle of radius $\ep_1$. $\ti R(z)$ is defined as shown, and $J$ is the
analytic continuation of the
jump matrix for $R$ on $\Si_R$. (It is easy to show that the continuation exists
in a neighborhood of the original $\Si_R$. Neighbourhoods where 2 lines intersect are 
analyzed similarly.)
$\ti R(z)$ satisfies the same Riemann-Hilbert problem as $R(z)$ but on
the contour $\ti\Si_R$.
The argument leading to (\ref{R2}) for $R$ and $\Si_R$ holds for $\ti R$ and
$\ti\Si_R$ as well. Therefore (see Figure 5),
\be\la{Restl}
|R(z)-I|=|\ti R(z)-I|\le c_3/n.
\ee
Analysis of the analytic continuation $J(z)$ shows that we can find the same
$c_3$ for all $\al_j$'s in a bounded set, $n>n_0$, and all
$z\in \bbc\setminus\Si_R$ up to the boundary $\Si_R$.
This is the estimate we need.

\begin{figure}
\centerline{\psfig{file=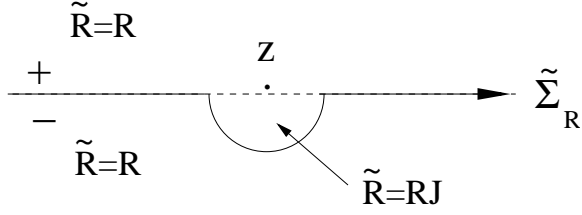,width=3.0in,angle=0}}
\vspace{0cm}
\caption{
Deformed part of the contour $\Si_R$.}
\label{fig5}
\end{figure}

We now proceed with the proof by induction.
On $\Si^\mathrm{out}$, we define $\De_j\equiv 0$.
Consider the function $R_1(z)$ analytic outside $\Si_R$, satisfying the
jump condition $R_{1+}(s)-R_{1-}(s)=\De_1(s)$, $s\in\Si_R$, and the condition
$R_{1}(z)=O(1/z)$ at infinity. The unique solution of this Riemann-Hilbert
problem is given by the Sokhotsky-Plemelj formula:
\be\la{R1l}
R_1(z)=C(\De_1)={1\over 2\pi i}\int_{\Si_R}\De_1(s){ds\over s-z}.
\ee
Because of the estimate
$\De_1(s)=O(1/n)$, uniform in
$\al_j$'s, $s\in\Si_R$ as $n\to\infty$, we have
\be\la{R1est}
R_1(z)=O(1/n),\qquad n\to\infty,
\ee
uniform in $\al_j$'s and $z$ satisfying ${\rm{dist}}(z,\Si_R)\ge\ep_1$.
Deforming the contour, we extend this estimate
to a uniform one for all $z\in\bbc\setminus\Si_R$.
By (\ref{R0}) and (\ref{R1l}),
\be
R(z)-I-R_1(z)=C(R_-\De-\De_1)=C((R_--I)\De+\De-\De_1).
\ee
The (uniform in $s\in\Si_R$ and $\al_j$'s) estimates
$R_-(s)-I=O(1/n)$ (\ref{Restl}), $\De(s)=O(1/n)$,
$\De(s)=O(\exp(-\ep n |s|))$ (on $\Si^{\mathrm{out}}$),
$\De(s)-\De_1(s)=O(1/n^2)$ (on $\partial U$) imply that
\be\la{R12est}
R(z)-I-R_1(z)=O(1/n^2),\qquad n\to\infty
\ee
uniformly in $\al_j$'s, and $z$ such that ${\rm{dist}}(z,\Si_R)\ge\ep_1$.
By a contour deformation argument this result extends uniformly for
$z\in\bbc\setminus\Si_R$.

Next solving the Riemann-Hilbert problem for $R_2$ (i.e., with the jump
$R_{2+}-R_{2-}=R_{1-}\De_1+\De_2$) we get
\be
R_2(z)=C(R_{1-}\De_1+\De_2).
\ee
By (\ref{R1est}) and the
estimates $\De_k(s)=O(1/n^k)$, we have as above for $R_1$:
\be
R_2(z)=O(1/n^2),\qquad n\to\infty,
\ee
uniformly in $\al_j$'s, $z\in\bbc\setminus\Si_R$.
By (\ref{Restl}), (\ref{R12est}), and the estimates for $\De$,
\be\eqalign{
R(z)-I-R_1(z)-R_2(z)=C(R_-\De-\De_1-R_{1-}\De_1-\De_2)=\\
C((R_--I)(\De-\De_1))+C(\De-\De_1-\De_2)+
C((R_--I-R_{1-})\De_1)=O(n^{-3}),\\
n\to\infty,}
\ee
with the same uniformity property in $\al_j$'s and $z$.

The general $k$'th induction step is carried out similarly. We have,
using estimates from the first $k-1$ steps,
\be
R_k(z)=C\left(\sum_{j=1}^{k}R_{k-j,-}\De_{j}\right)=
O(n^{-k}).
\ee
and for the error after $k$ terms
\be\eqalign{
R(z)-I-\sum_{j=1}^k R_k(z)=
C\left(R_-\De-\sum_{j=1}^k\sum_{i=1}^j R_{j-i,-}\De_i\right)=\\
C\left((R_--I)\left[\De-\sum_{j=1}^{k-1}\De_j\right]\right)+
C\left(\De-\left[\sum_{j=1}^k\De_j\right]\right)+\\
C\left(\sum_{j=1}^{k-1}\left[R_--I-\sum_{i=1}^{k-j}R_{i-}\right]\De_j\right)=
O(n^{-k-1}).}\la{104}
\ee
The last 2 estimates are valid uniformly
for $\al_j$'s in a bounded set and $z\in\bbc\setminus\Si_R$.
This concludes the proof of (\ref{Ras}--\ref{plem}).

%%%%%%%%%%%%%%%%%%%%%%%%%%%%%%%%%%%%%%%%%%%%%%%

\subsection{Calculation of the asymptotics for the polynomials}

In the intersection of the region A (the area outside the lenses)
and the neighborhoods $U$ we have by (\ref{Rb}), (\ref{defS}), (\ref{Ras}),
\be\la{last}
T(z)=S(z)=P_\theta(z)+R_1(z)P_\theta(z)+O(1/n^2)P_\theta(z),\qquad 
z\in U_\theta\cap A,\qquad 
\theta=\pm1,\lb_j,\infty.
\ee

It turns out that to prove Theorem 1 using (\ref{id}), we need to
evaluate only $P_{\lb_j}(z)$ as $z\to\lb_j$ (i.e., the main asymptotic
term of $T$ at $\lb_j$) and (\ref{last}) as $z\to\infty$
(i.e., the first 2 terms of $T$ at infinity).

We evaluate $P_{\lb_j}(z)$ as $z\to\lb_j$ inside the region $z(II)$ (see Figure 3).
$P_{\lb_j}(z)$ there is given by (\ref{Plb}), and (\ref{Psi}).

Tracing back the transformations (\ref{last}), (\ref{T}), and (\ref{U}), 
we obtain for the main asymptotic term 
\be\la{Ylb}
Y(z\sqrt{2n})=(2n)^{n\si_3/2}e^{nl\si_3/2}
(2n)^{\A\si_3/2}
(I+O(1/n))P_{\lb_j}(z)(2n)^{-\A\si_3/2}
e^{n(g(z)-l/2)\si_3},
\qquad z\in z(II).
\ee

For application in the identity (\ref{id}), we need to analyze equation
(\ref{Ylb}) as $z$ approaches $\lb_j$ along a path in $z(II)$. 
Note first (recall (\ref{Dom})) that
\be\eqalign{
{\cal D}(z)W_{j}(z)^{-1}=
e^{i\pi\ti\al_j}(z+\sqrt{z^2-1})^{-\A}=
e^{i\pi\ti\al_j-i\A(\pi/2-\tau_j)}(1+O(z-\lb_j)),\\
{1\over\sqrt{2}}(\sqrt{\lb_j+1}+{\sqrt{\lb_j-1}}_+)=e^{i(\pi/2-\tau_j)/2},\qquad
\tau_j=\arcsin\lb_j\qquad \ti\al_j=\sum_{k=j}^m\al_k.}
\ee
Therefore, the product of the first 3 factors in (\ref{Plb}) gives
as $z\to\lb_j$:
\be\eqalign{
\frac{2^{-\A\si_3}}{\sqrt{2}(1-\lb_j^2)^{1/4}}
\pmatrix{e^{-(\A+1/2)\tau_j i -(2\ti\al_j-\A-\al_j)i\pi/2+n\phi_+(\lb_j)}&
e^{(\A+1/2)\tau_j i+(2\ti\al_j-\A-\al_j)i\pi/2 -n\phi_+(\lb_j)}\cr
-e^{-(\A-1/2)\tau_j i-(2\ti\al_j-\A-\al_j)i\pi/2 +n\phi_+(\lb_j)}&
e^{(\A-1/2)\tau_j i+(2\ti\al_j-\A-\al_j)i\pi/2 -n\phi_+(\lb_j)}}\times\\
(1+O(z-\lb_j)).}
\ee
By (\ref{zlb}), we need the asymptotics of Bessel functions at a small argument 
for (\ref{Psi}) (see \ci{Abr} for them). For $\al_j\neq 1/2+k$, $k=0,1,\dots$,
\be\eqalign{
\Psi_{\al_j}(\ze)=\\
\pmatrix{
C_1(\al_j)\ze^{\al_j+1}&
-{1\over\sqrt{2\pi}}
(\ze/2)^{-\al_j}e^{i{\pi\over 2}(2\al_j+1/2)}\Gamma(\al_j+1/2)+C_2(\al_j)\ze^{\al_j+1}\cr
-i(\ze/2)^{\al_j}e^{-i{\pi\over 2}(2\al_j-1/2)}{\sqrt{2\pi}\over\Gamma(\al_j+1/2)}&
C_3(\al_j)\ze^{-\al_j+1}+C_4(\al_j)\ze^{\al_j}}\times\\
(1+O(\ze)),\qquad z\in z(II),}
\ee
where $C_j(\al_j)$ are constants whose precise expressions will not be important below.
For $\al_j= 1/2+k$, $k=0,1,\dots$, the constants $C_1$ and $C_3$ remain the same, while
$C_2(\al_j)$ is replaced with $C_5(\al_j)+C_6(\al_j)\ln\ze$, and 
$C_4(\al_j)$ is replaced with $C_7(\al_j)+C_8(\al_j)\ln\ze$.

Now the part of the product of the next 5 factors 
in (\ref{Plb}) with the $\ze^{2\al_j}$ terms omitted
(``regularized'' part)
gives for $z\to\lb_j$ and $\al_j\neq0$:
\be\la{psireg}\eqalign{
\left(e^{-i\pi\si_3/4}{1\over\sqrt{2}}\pmatrix{1& i\cr i& 1}
\Psi_{\al_j} (n\hat f(z))W_{j}(z)^{-\si_3}\right)_{\mathrm reg}\to\\
\pmatrix{1& -{1\over 2}\cr
1& {1\over 2}}
\left({\sqrt{\pi}\left(n\sqrt{1-\lb_j^2}\right)^{\al_j}\over
    \Gamma(\al_j+1/2)}\right)^{\si_3}\prod_{k\neq j}|\lb_k-\lb_j|^{-\al_k\si_3}.}
\ee
This result will correspond to the ``principal value'' part of $Y$ given by (\ref{reg}). 
The omitted part of order $\ze^{2\al_j}$ 
does not contribute to the ``principal value'' part of the total product for $Y$
because all the correction terms are $O(\ze)$ as $\ze\to 0$ and $\Re\al_j>-1/2$. 

The expression (\ref{psireg}) is also valid for $\al_j=0$ (since, e.g., 
(\ref{Sing}) for this case can be obtained by letting $\al_j\to 0$ in 
the $\ze^{2\al_j}$ term with $\al_j<0$).

Finally, the product of the last factors in (\ref{Plb}) and (\ref{Ylb}) gives 
by (\ref{g}):
\be
e^{n(g(z)-\phi(z)-l/2)}=
e^{n(g_+(\lb_j)-\phi_+(\lb_j)-l/2)}(1+O(z-\lb_j))=
e^{n\lb_j^2\si_3}(1+O(z-\lb_j)).
\ee
Taking the product of all contributions, we have:
\begin{eqnarray}
&\eqalign{
\lim_{z\to\lb_j}Y^{\rm vp}(z\sqrt{2n})=
\left({n\over2e}\right)^{n\si_3/2}
(2n)^{\A\si_3/2}
(I+O(1/n))
2^{-\A\si_3}
(1-\lb_j^2)^{-1/4}\times\\
\pmatrix{
\sqrt{2}\cos {1\over 2}(t_j-\tau_j)&
-{i\over\sqrt{2}}\sin{1\over 2}(t_j-\tau_j) \cr
-i\sqrt{2}\sin{1\over 2}(t_j+\tau_j) &
{1\over\sqrt{2}}\cos{1\over 2}(t_j+\tau_j) }
\left({\sqrt{\pi}\left(n\sqrt{1-\lb_j^2}\right)^{\al_j} e^{n\lb_j^2}\over
    (2n)^{\A/2}\Gamma(\al_j+1/2)}\right)^{\si_3}
\prod_{k\neq j}|\lb_k-\lb_j|^{-\al_k\si_3},}\la{asY}\\
& t_j=2\pi
  n\int_{\lb_j}^1\psi(y)dy+\pi\al_j-2\pi\sum_{i=j}^m\al_i
+\A(\pi-2\tau_j),\qquad \tau_j=\arcsin\lb_j.\la{t}
\end{eqnarray}

We now turn to the asymptotics of the coefficients $\ka_n$, $\beta_n$ and
$\ga_n$ of $p_n(z)$. Here we need the first 2 asymptotic terms in $n$. As usual,
we compute them investigating the limit $z\to\infty$ of $Y(z)$. 
Namely, by (\ref{RHM}),
\be\la{YU1}
\ka^2_{n-1}=\lim_{z\to\infty}{iY_{21}(z)\over 2\pi z^{n-1}},\qquad
U_{11}(z)=z^n+{\beta_n\over\sqrt{2n}}z^{n-1}+
{\ga_n\over 2n}z^{n-2}+\cdots
\ee
As $z\to\infty$, we need to know asymptotics of $Y(z)$ in the region
$A$, which are given by the expressions (cf. (\ref{Ylb})):
\be\eqalign{
Y(z\sqrt{2n})=(2n)^{n\si_3/2}U(z),\\
U(z)=e^{nl\si_3/2}(2n)^{\A\si_3/2}(I+R_1(z)+O(1/n^2))P_\infty(z)
e^{n(g(z)-l/2)\si_3}(2n)^{-\A\si_3/2},\\
z\in A\cap U_\infty.}\la{YU2}
\ee
Let us compute $R_1(z)$  using (\ref{plem}). Consider first the 
neighborhood $U_{\lb_j}$.
Substituting $\De_1(x)$ given by (\ref{alb}) into (\ref{plem}) and calculating
residues at a simple pole $x=\lb_j$, we obtain the contribution to $R_1$
from the neighborhood $U_{\lb_j}$:
\be\eqalign{
R_1^{(\lb_j)}(z)=
{1\over 2\pi i}\int_{\partial U_{\lb_j}}{\De_1 dx\over x-z}=
-{1\over z}\left(1+{\lb_j\over z}+O(z^{-2})\right)
{1\over 2\pi i}\int_{\partial U_{\lb_j}}\De_1 dx=\\
{1\over z}\left(1+{\lb_j\over z}+O(z^{-2})\right)
{{\cal D}_\infty^{\si_3}\al_j\over 4n(1-\lb_j^2)}
\pmatrix{-\al_j\lb_j+\sin t_j& i(\al_j-\cos(t_j-\tau_j))\cr
i(\al_j+\cos(t_j+\tau_j))& \al_j\lb_j-\sin t_j}{\cal D}_\infty^{-\si_3},}\la{R1lb}
\ee
where $t_j$, $\tau_j$ are defined in (\ref{t}).

To compute the contribution from the neighborhood $U_{1}$, 
we repeat the calculation now using $\De_1(z)$ from (\ref{a1}).
An additional
complication is that we need to calculate residues in the pole $z=1$ of
second order. We obtain:
\be\eqalign{
R_1^{(1)}= {1\over 2\pi i}\int_{\partial U_1}{\De_1 dx\over x-z}=
{1\over z}\left(1+{1\over z}+O(z^{-2})\right)
{{\cal D}_\infty^{\si_3}\over 8n}
\pmatrix{1/8-\A^2& i/6+i\A+i\A^2\cr
         i/6-i\A+i\A^2& -1/8+\A^2}{\cal D}_\infty^{-\si_3}+\\
{1\over z^2}{5{\cal D}_\infty^{\si_3}\over 8\cdot 24n}
\pmatrix{-1&i\cr i&1}{\cal D}_\infty^{-\si_3}.}\la{R11}
\ee
Here the dependence on $\al_j$ comes from the expansion (see (\ref{Dom})):
\be\la{Dom1}
{{\cal D}^2(z)\over\om(z)}=
1-2\sqrt{2}\A\sqrt{z-1}+4\A^2(z-1)+O((z-1)^{3/2}),\qquad\mbox{as
  $z\to 1$.}
\ee

A similar calculation for $U_{-1}$ gives
\be\eqalign{
R_1^{(-1)}= {1\over 2\pi i}\int_{\partial U_{-1}}{\De_1 dx\over x-z}=
{1\over z}\left(1-{1\over z}+O(z^{-2})\right)
{{\cal D}_\infty^{\si_3}\over 8n}
\pmatrix{-1/8+\A^2& i/6+i\A+i\A^2\cr
         i/6-i\A+i\A^2& 1/8-\A^2}{\cal D}_\infty^{-\si_3}+\\
{1\over z^2}{5{\cal D}_\infty^{\si_3}\over 8\cdot 24n}
\pmatrix{-1&-i\cr -i&1}{\cal D}_\infty^{-\si_3}.}\la{R1-1}
\ee
In this case
\be\la{Domm1}
{{\cal D}^2(z)\over\om(z)}=
1+2\sqrt{2}i\A\sqrt{z+1}-4\A^2(z+1)+O((z+1)^{3/2}),\qquad\mbox{as
  $z\to -1$.}
\ee

Summing up all the contributions (\ref{R1lb}), (\ref{R11}), 
and (\ref{R1-1}), we obtain:
\be
R_1=R_1^{(1)}+R_1^{(-1)}+\sum_{j=1}^m R_1^{(\lb_j)}.
\ee
Substituting this into (\ref{YU2}) and using the expansions for $z\to\infty$:
\be\eqalign{
{{\cal D}_\infty \over {\cal D}(z)}
=1+{1\over z}\sum_{j=1}^m\al_j\lb_j+
{1\over 2z^2}\left[\left(\sum_{j=1}^m\al_j\lb_j\right)^2 +
\sum_{j=1}^m\al_j\lb_j^2-{1\over 2}\A\right]+O(z^{-3}),\\
a(z)=1-{1\over 2z}+{1\over 8z^2}+O(z^{-3}),\qquad
g(z)=\ln z - {1\over 8z^2}+O(z^{-4}),}\la{auxexp}
\ee
we finally obtain from (\ref{YU1})
\be
\ka_{n-1}^2={2^{n-1+\A}n^{-\A}\over\sqrt{\pi}(n-1)!}\left\{
1-{1\over 2n}\left(\A^2-\A+\sum_{j=1}^m
{1\over 1-\lb_j^2}\left[\al_j^2+\al_j\cos(t_j+\tau_j)\right]\right)+O\left({1\over
  n^2}\right)\right\},\la{k-as}
\ee
\be
\beta_n=\sqrt{2n}\left\{\sum_{j=1}^m\al_j\lb_j+
{1\over 4n}\sum_{j=1}^m{\al_j\sin t_j -\al_j^2\lb_j \over 1-\lb_j^2}+
O\left({1\over  n^2}\right)
\right\},\la{b-as}
\ee
\be
\eqalign{
\ga_n=n\left\{-{n-1\over 4}+
\left(\sum_{j=1}^m\al_j\lb_j\right)^2+\sum_{j=1}^m\al_j\lb_j^2-{\A\over 2}
+{1\over 4n}\left[\A-\A^2+\sum_{j=1}^m\al_j^2\right.\right.\\
\left.\left.
-\sum_{j=1}^m{\al_j\cos(t_j+\tau_j) +\al_j^2\lb_j^2 \over 1-\lb_j^2}+
2\sum_{j=1}^m{\al_j\sin t_j -\al_j^2\lb_j \over 1-\lb_j^2}\sum_{k=1}^m\al_k\lb_k
\right]+O\left({1\over  n^2}\right)\right\},}\la{g-as}
\ee
where $\tau_j=\arcsin\lb_j$, $\A=\sum_{j=1}^m\al_j$, 
and $t_j$ is defined by (\ref{t}) and (\ref{gf}).
The error terms here and in (\ref{asY}) are uniform 
for $\al_j$ in a bounded set provided only $\Re\al_j>-1/2$.
Let us track the $\al$-dependence of the error terms in more detail. 
These terms in (\ref{k-as})--(\ref{g-as}) are those in (\ref{YU2}), at worst 
multiplied by a polynomial in $\al_j$ (independent of $n$) and $2^{-\A}$
coming from the expansion (\ref{auxexp}) of $D(z)$. 
The error term in (\ref{asY}) is the same as the one in (\ref{Ylb}).
Now the error terms in (\ref{Ylb}) and (\ref{YU2}) are those from (\ref{Ras})
which we showed to have the above uniformity property.

Thus we constructed a solution to the Riemann-Hilbert problem of Section 2 for $n>n_0$,
$\al_j$ in any bounded set of the half-plane
$\Re\al_j>-1/2$. By uniqueness, it gives the orthogonal polynomials via (\ref{RHM}).
On the other hand, the determinantal representation for the orthogonal polynomials
shows that $R(z)$ is an analytic function of $\al_j$'s. 
Furthermore, $R_k(z)$ are
also analytic functions of $\al_j$'s by construction. 
Thus, the error term in (\ref{104})
is both analytic and uniform in $\al_j$'s in a bounded set of the half-plane
$\Re\al_j>-1/2$. Therefore, it is differentiable in $\al_j$'s (the derivative being
 of the same order in $n$ and uniform in $\al_j$'s). 
Hence, we easily conclude that the error terms in 
(\ref{asY}), (\ref{k-as})--(\ref{g-as}) have the same differentiability property.
Alternatively, we could have deduced the differentiability of the error terms by 
noticing first that the asymptotic expansions of Bessel functions we used 
are differentiable in $\al_j$'s.

%%%%%%%%%%%%%%%%%%%%%%%%%%%%%%%%%%%%%%%
%%%%%%%%%%%%%%%%%%%%%%%%%%%%%%%%%%%%%%%
%%%%%%%%%%%%%%%%%%%%%%%%%%%%%%%%%%%%%%%
%%%%%%%%%%%%%%%%%%%%%%%%%%%%%%%%%%%%%%%
%%%%%%%%%%%%%%%%%%%%%%%%%%%%%%%%%%%%%%%
%%%%%%%%%%%%%%%%%%%%%%%%%%%%%%%%%%%%%%%
%%%%%%%%%%%%%%%%%%%%%%%%%%%%%%%%%%%%%%%
%%%%%%%%%%%%%%%%%%%%%%%%%%%%%%%%%%%%%%%
%%%%%%%%%%%%%%%%%%%%%%%%%%%%%%%%%%%%%%%
%%%%%%%%%%%%%%%%%%%%%%%%%%%%%%%%%%%%%%%
%%%%%%%%%%%%%%%%%%%%%%%%%%%%%%%%%%%%%%%
%%%%%%%%%%%%%%%%%%%%%%%%%%%%%%%%%%%%%%%

\section{Proof of Theorem 1}
We now substitute the asymptotics (\ref{k-as})--(\ref{g-as}), and (\ref{asY}) into the
differential identity (\ref{id}). We assume first that all 
$\{\al_j\}_{j=1}^m\in \wt\Om\setminus\hat\Om$ (see Section 2) and $\{0\}\in\wt\Om$.
Care is needed with estimation of
$\ka_n$. To obtain the 
asymptotics of $\ka_n$ from (\ref{k-as}) we need first to replace 
$n$ with $n+1$ and second, to replace $\lb_j$ with $\lb_j\sqrt{n\over n+1}$.
Without this second step we would obtain a coefficient corresponding
to the weight $\prod_j|x-\lb_j\sqrt{2(n+1)}|^{2\al_j}e^{-x^2}$. However, since
the new $\lb_j$ is inside $(-1,1)$ for $n$ large enough, and because it
enters the asymptotics of $\ka_n$ starting with $O(1/n)$ term, this
second replacement affects only $t_j$ in (\ref{k-as}). 
We obtain
\be\eqalign{
-(n+2\sum_{j=1}^m\al_j)(\ln\ka_n\ka_{n-1})'_{\al_\nu}-
2\left(\ka_{n-1}\over \ka_n\right)^2
\left(\ln{\ka_{n-1}\over\ka_n}\right)'_{\al_\nu}+\\
2\left[\ga'_{n,\al_\nu}-\beta_n\beta'_{n,\al_\nu}\right]=
(n+2\sum_{j=1}^m\al_j)\ln(n/2)+
(2\lb_\nu^2-1)n+2\al_\nu+O(1/n).}\la{55}
\ee
Furthermore,
\be\eqalign{
2\sum_{j=1}^m\al_j\left(Y_{11}(\mu_j)'_{\al_\nu} Y^{\rm vp}_{22}(\mu_j)-
Y_{21}(\mu_j)'_{\al_\nu} Y^{\rm vp}_{12}(\mu_j)
+(\ln\ka_n\ka_{n-1})'_{\al_\nu}Y_{11}(\mu_j) Y^{\rm vp}_{22}(\mu_j)\right)=\\
\al_\nu\ln(1-\lb_\nu^2)-2\al_\nu{\Gamma'(\al_\nu+1/2)\over\Gamma(\al_\nu+1/2)}
-2\sum_{j\neq\nu}\al_j\ln(n|\lb_j-\lb_\nu|)
+O\left({\ln n\over n}\right).}\la{56}
\ee
Note that the trigonometric terms of (\ref{k-as})--(\ref{g-as}), and (\ref{asY})
cancel in these expressions. 

Legitimacy of differentiation of the error terms 
w.r.t. $\al$ (and uniformity of these terms and their derivatives) follows 
from that for 
the error terms in (\ref{k-as})--(\ref{g-as}), and (\ref{asY})
explained in the end of the previous section.

The sum of (\ref{55}) and (\ref{56}) yields:
\be\la{idfin}\eqalign{
{d\over d\al_\nu}\ln D_n(\al_1,\al_2,\dots,\al_m)=
(n+2\al_\nu)\ln(n/2)+
(2\lb_\nu^2-1)n+2\al_\nu+
\al_\nu\ln(1-\lb_\nu^2)-\\
2\al_\nu{\Gamma'(\al_\nu+1/2)\over\Gamma(\al_\nu+1/2)}
-2\sum_{j\neq\nu}\al_j\ln(2|\lb_j-\lb_\nu|)
+O\left({\ln n\over n}\right),\qquad \nu=1,2,\dots,m.}
\ee

Since the error term here is uniform in all $\al_\nu$, we can integrate 
this identity.

First, set $\al_2=\cdots=\al_m=0$, $\nu=1$ in (\ref{idfin}).
This identity was established for $\al_1$ outside the set $\Om(*,0,\dots,0)$
(see Section 2).
Note that the number of points in $\Om(*,0,\dots,0)$ is at most finite.
Indeed, the function
$\ka_k^2=\ka_k^2(\al_1,0,\dots,0)$ is a ratio $D_k/D_{k+1}$ of two analytic functions 
of $\al_1$ for $\Re\al_1>-1/2$. They are not identically zero because they are 
known to be positive for $\al_1=0$. Consider the function 
$f(\al_1)=D_n(\al_1,0,\dots,0)\exp(-\int_0^{\al_1} r(n,z)dz)$, 
where $r(n,\al_1)$ is the r.h.s.
of (\ref{idfin}) with $\nu=1$, $\al_2=\cdots=\al_m=0$. 
Equation (\ref{idfin}) is the statement
that $f'(\al_1)=0$ outside $\Om(*,0,\dots,0)$. Since $f(\al_1)$ is continuously 
differentiable, $f'(\al_1)=0$ for {\it all} $\al_1$ such that 
$(\al_1,0,\dots,0)\in\wt\Om$.
Moreover, as $f(\al_1)=f(0)=D_n(0,\dots,0)\neq 0$, 
the determinant $D_n(\al_1,0,\dots,0)$ is 
never zero. Hence the identity (\ref{idfin}) with $\nu=1$, $\al_2=\cdots=\al_m=0$
holds for {\it all} $\al_1$ such that $(\al_1,0,\dots,0)\in\wt\Om$. 
Integrating it over $\al_1$ from $0$ to some $\al_1$, we obtain:
\be\la{fin1}\eqalign{
\ln D_n(\al_1,0,\dots,0)=
\ln C(\al_1)+{\al_1^2\over 2}\ln(1-\lb_1^2)+
(\al_1 n +\al_1^2)\ln{n\over 2}+n(2\lb_1^2-1)\al_1+\\
\ln D_n(0,\dots,0)+O\left({\ln n\over n}\right),}
\ee
where
\be
C(\al)=\Gamma(\al+1/2)^{-2\al}\exp\left(2\int_0^\al
  \ln\Gamma(s+1/2)ds+
\al^2\right).
\ee
By freedom in the choice of $\wt\Om$, the expansion (\ref{fin1}) is valid for 
any $\al_1$, $\Re\al_1>-1/2$.

Second, a reasoning similar to the one above shows that equation (\ref{idfin})
with $\al_1$ fixed and $\al_3=\cdots=\al_m=0$ holds for all $\al_2$, 
$\Re\al_2>-1/2$.
Integrating it over $\al_2$ from $0$ to some $\al_2$ gives
\be\la{fin2}\eqalign{
\ln D_n(\al_1,\al_2,0,\dots,0)=
\ln C(\al_2)+{\al_2^2\over 2}\ln(1-\lb_2^2)+
(\al_2 n +\al_2^2)\ln{n\over 2}+n(2\lb_2^2-1)\al_2\\
-2\al_1\al_2\ln 2|\lb_1-\lb_2|+
\ln D_n(\al_1,0,\dots,0)+O\left({\ln n\over n}\right)=\\
\ln C(\al_1)C(\al_2) +\sum_{j=1}^2{\al_j^2\over 2}\ln(1-\lb_j^2)+
\sum_{j=1}^2(\al_j n +\al_j^2)\ln{n\over 2}+
n\sum_{j=1}^2\al_j(2\lb_j^2-1)\\
-2\al_1\al_2\ln 2|\lb_1-\lb_2|+
\ln D_n(0,\dots,0)+O\left({\ln n\over n}\right).}
\ee
To obtain the second equation, we substituted $\ln D_n(\al_1,0,\dots,0)$
from (\ref{fin1}). Equation
(\ref{fin2}) is valid for any $\al_1$, $\al_2$, $\Re\al_1,\Re\al_2>-1/2$.

Proceeding in this way, we prove Theorem 1 by induction
after $m$ steps. 

The second representation (\ref{c}) for the constant $C(\al)$ is easy to
obtain using the following properties of the G-function \ci{WW}:
\be\eqalign{
G(z+1)=\Gamma(z)G(z),\qquad G(1)=1,\\
 \int_0^z\ln \Gamma(x+1)dx=
{z\over 2}\ln 2\pi -{z(z+1)\over2}+z\ln\Gamma(z+1)-\ln G(z+1).}
\ee
Another identity is useful for comparison with (\ref{D0}):
\be
2\ln G(1/2)=
(1/12)\ln2-\ln\sqrt\pi+3\zeta'(-1).
\ee

\section{Acknowledgements}
I thank Y. Fyodorov (from whom I learned about this problem) and A. Kuijlaars 
for useful discussions and comments. 
I am very grateful to P. Deift and A. Its for many important 
suggestions, and to V. Kravtsov and O. Yevtushenko for their 
hospitality at the Abdus Salam ICTP where this work was started.

\end{document}